# Adaptive Transit Design: Optimizing Fixed and Demand Responsive Multi-Modal Transportation via Continuous Approximation

Giovanni Calabrò[a], Andrea Araldo[b], Simon Oh[c], Ravi Seshadri[d], Giuseppe Inturri[e], Moshe Ben-Akiva[f]

[a] University of Catania, Via Santa Sofia, 64, 95123 Catania, Italy; giovanni.calabro@unict.it
[b] Samovar, Télécom SudParis, Institut Polytechinque de Paris, 19 place Marguerite Perey, 91120 Palaiseau, France; andrea.araldo@telecom-sudparis.eu
[c] Department of Autonomous Mobility, Korea University, Sejong 30019, Korea; simonoh@korea.ac.kr
[d] Department of Technology, Management and Economics, Technical University of Denmark, 2800 Kongens Lyngby, Denmark; ravse@dtu.dk
[e] Department of Electrical, Electronic and Computer Engineering, University of Catania, Via Santa Sofia, 64, 95123 Catania, Italy; giuseppe.inturri@unict.it
[f] Department of Civil and Environmental Engineering, Massachusetts Institute of Technology, Cambridge, MA 02139, United States; mba@mit.edu

Abstract

In most cities, transit consists solely of fixed-route transportation, whence the inherent limited Quality of Service for travellers in suburban areas and during off-peak periods. On the other hand, completely replacing fixed-route (FR) with demand-responsive (DR) transit would imply a huge operational cost. It is still unclear how to integrate DR transportation into current transit systems to take full advantage of it. We propose a Continuous Approximation model of a transit system that gets the best from fixed-route and DR transportation. Our model allows deciding whether to deploy a FR or a DR feeder, in each sub-region of an urban conurbation and each time of day, and to redesign the line frequencies and the stop spacing of the main trunk service. Since such a transit design can adapt to the spatial and temporal variation of the demand, we call it *Adaptive Transit*. Numerical results show that, with respect to conventional transit, *Adaptive Transit* significantly improves user-related cost, by drastically reducing access time to the main trunk service. Such benefits are particularly remarkable in the suburbs. Moreover, the generalized cost, including agency and user cost, is also reduced. These findings are also confirmed in scenarios with automated vehicles. Our model can assist in planning future-generation transit systems, able to improve urban mobility by appropriately combining fixed and DR transportation.

## 1. Introduction

In recent years, urban transportation has witnessed the birth and spread of new demand-responsive and ride-hailing services, mostly provided by private companies (e.g., Uber, Lyft, Via), which connect community drivers with passengers via mobile applications. In most cities, these "user-centric" services have penalized the conventional public transit (PT), which has basically not evolved in the last decades and still consists in fixed routes and fixed scheduling with some exceptions, or pilots or services for a specific targeted population (elders or handicapped). However, the detrimental role of ride-hailing and ride-sourcing services (i.e., transport services connecting community drivers with passengers via mobile applications) towards PT can be reversed by transforming them from PT substitutes to PT complement (Sadowsky and Nelson, 2017). In order to pursue this transformation, it is essential to rethink the whole transit network via a multi-modal approach and an integrated design of the various transportation modes (conventional and demand-responsive).

Conventional PT is inefficient in sparse demand areas, i.e., where few trip requests per $km^2$ occur. Indeed, providing a high number of lines with an adequate frequency to ensure an acceptable Quality of Service (QoS) to travellers would result in low passenger load factors and thus in an excessive agency-related cost. This problem is evident in suburbs and is one of the reasons for geographical inequity (Giuffrida et al., 2017; Badeanlou et al., 2022). On the other hand, demand-responsive transportation is not the solution to all mobility needs, as it is not suitable to serve dense demand (Basu et al., 2018), as it would result in tortuous vehicle routes (Araldo, Gao et al., 2019), high operational cost and poor QoS.

Therefore, a combination of fixed-route (FR) and demand-responsive (DR) services may guarantee high capacity for dense demand areas and, at the same time, acceptable QoS in sparse demand areas (Calabrò et al., 2022). For these reasons, in recent years public authorities have launched pilots to experiment with different ways to complement their offer with on-demand services, by subsidizing ride-sharing companies (McCoy et al., 2018 and Sörensen et Al, 2021). Moreover, the scientific community has done a big effort in modelling the performance of DR transportation. However, there is no systematic methodology to guide the design of future-generation transit systems integrating FR and DR transportation.

The contribution of this paper, toward filling this gap, can be summarized as follows:

- We are the first to propose, to the best of our knowledge, a Continuous Approximation model and an optimization procedure to design a transit system combining both FR and DR

- transportation. We call such a system *Adaptive Transit*. It consists of mass rapid transit (MRT), which is always FR, and a feeder service provided by bus. Depending on the sub-region of the conurbation and period of the day, the system changes the feeder operation, between FR and DR, in order to adapt to the spatial and temporal variation of the demand density. The optimization procedure decides the deployment parameters of both MRT and feeder (frequencies, stop spacing, etc.).
- Via extensive numerical evaluation, we compare *Adaptive Transit* with the conventional transit design, where the feeder service is always FR. Our results show that *Adaptive Transit* improves the user QoS, in particular during off-peak hours and in suburban areas, while keeping the overall cost (which includes both the agency-related cost and the user-related cost) under a reasonable level, even slightly reducing it.

The main novelty of this work is the *joint* optimization of both FR and DR transportation, integrated in a single system, deciding the overall deployment over space and in different times of day. The paper is structured as follows: we discuss the related work in Section 2, present the considered design schemes of transit in Section 3, among which the *Adaptive Transit*. We then present a Continuous Approximation model of such schemes and the procedure to compute their optimal structure (Section 4). We finally contrast the performance of *Adaptive Transit* with conventional transit schemes in numerical results (Section 5) and conclude the paper (Section 6). Additionally, we test the benefits of *Adaptive Transit* in a small-scale scenario with agent-based simulation (Appendix B). For the sake of reproducibility, we release the Matlab code of the CA model and the design optimization procedure as opensource.[1]

Table A1 reports the acronyms used throughout the paper.

## 2. Related work

During the last decades, transit network design has been studied via several optimization problems, based on different objectives (user and/or operator cost minimization, total welfare maximization, protection of the environment, etc.), parameters and decision variables (network structure, demand patterns, fleet characteristics, headway, route and stop spacing, etc.) and solution methodologies (analytical, heuristics or meta-heuristics). An extensive review on Transit Route Network Design Problems is provided by Farahani et al. (2013).

---

[1] https://github.com/giovanni-cal/future-transit

Only in the last decade, the concept of Mobility as a Service (MaaS) has emerged, in which different modes of transportation are integrated into a single multi-modal offer (Smith et al., 2018; Le Pira et al., 2021). We believe that MaaS should not consist in just adding DR services on top of the existing FR services. On the contrary, to get the most benefits from a multi-modal transit combining the two, it is required to holistically redesign the entire transit and "co-design" FR and DR services. Classic transit design methods are not suitable to this new aim, and new approaches are needed.

In this section, we first briefly motivate our choice of a feeder-trunk structure for transit, resorting to the literature showing its advantages (Section 2.1). We then introduce the work on Continuous Approximation (CA), which is the modelling approach we adopt in this paper, with a particular focus on studies combining or comparing FR and DR operation (Section 2.2). We also discuss the work combining FR and DR transportation with methods different than CA (Section 2.3). We finally state the novelty of our work with respect to the state-of-the-art (Sectieon 2.4).

## 2.1 Feeder-trunk transit structure

The so-called "weak demand areas" (i.e., areas with low residential density and high motorization rate) are the most critical for conventional public transit, which is unable to ensure at the same time coverage, ridership and cost-efficiency. In these cases, an effective design of fixed-route feeder (FRF) or demand-responsive feeder (DRF) bus lines connecting weak demand areas with MRT nodes could therefore help to shift passenger's mode of transportation from individual to collective mobility (Calabrò et al., 2020a; 2020b; 2022), thus enhancing the accessibility to urban facilities and services.

The advantages of mass transit corridors in the metropolitan transportation supply have been shown by Mohaymany and Gholami (2010) and Gschwender et al. (2016). In particular, Gschwender et al. (2016) compared the feeder-trunk scheme against different direct lines structures (where no transfers are required) showing that the first structure performs better when the demand is quite low and dispersed and the distances to travel are high (results are however sensitive to the penalty value assigned to transfers). Mohaymany and Gholami (2010) demonstrated that feeder lines increase the use of high-capacity mass transit because the travel demand for a more extended area can be satisfied. Lau and Susilawati (2021) obtained similar findings in a simulation-based study concerning the impact of automated vehicles with "predefined routes", acting as a bus feeder line.

## 2.2 Continuous Approximation models in transit-related studies

Addressing transit network design problems at a strategic level with detailed models is often unfeasible when dealing with large-scale instances, as, in addition to their computational burden, such models are not robust to stochasticity and uncertainty of input data (Daganzo, 1987).

To overcome such limitations, Continuous Approximation (CA) models have been proposed. We comply with the literature Ansari et al. (2018) and define a CA model as the one where demand and supply variables, either as input or as decision variables, are continuous density functions over space. Such models are simple but powerful tools for the strategic stage of a transit plan. The key idea, as reported by Ansari et al. (2018), is to construct an objective function, including agency- and user-centric costs, based on the integration of localized functions of x-y coordinates, which can be analytically optimized without huge computational efforts. Results obtained via CA provide general insights about the performance of a whole transit system. It must be noted however that CA models are approximated and lack realism. For instance, they cannot include details about transit network topologies, traveller behaviour and vehicle routing. However, the CA methodology can provide useful insights in understanding, at high level, the impact of different design choices on the performance of a transportation network.

### 2.2.1 Continuous Approximation for Demand-Responsive transportation

Analytical models for demand-responsive transportation under a many-to-many demand pattern were proposed in Daganzo (1978)[2] for door-to-door services. CA models considering checkpoints, around which the demand can be clustered, were presented in Daganzo (1984) and Quadrifoglio et al. (2006).

Further work focused on DR transportation to serve the First Mile/Last Mile (FMLM), in particular comparing the performance of FRF and DRF therein. On this account, Quadrifoglio and Li (2009, 2010) and Papanikolau and Basbas (2021) estimated the demand density threshold, for a feeder transit service, below which DRF operations are more efficient than FRF ones. Edwards and Watkins (2013) expanded the comparison to a broader range of street networks, transit schedules and passenger demand levels. Recently, Badia and Jenelius (2020, 2021) found via CA that electrification and automation will impact the cost structure, so that the situations in which DR will be preferable to FR will extend (although FR will still be irreplaceable in very high demand-density areas).

---

[2] We include it in this subsection although it is not strictly a CA model

Note that, while the work mentioned in this subsection only focused on single FMLM sub-regions, we instead aim to devise a design for an entire urban area, consisting of many FMLM sub-regions.

### 2.2.2 Continuous Approximation for metropolitan-scale transit

Daganzo (2010) proposed a CA model of a FR transit network with a "hybrid" structure, which combines the advantage of both the grid (double transit routes coverage in the central area) and the hub-and-spoke (radial routes branching to the periphery) structures. Transit is described by only three decision variables: stop spacing, vehicle headway and ratio between the side of the central area (enjoying better coverage) and the side of the city boundary. The author found that the more expensive the system's infrastructure, the more it should tilt toward the hub-and-spoke concept. In all cases, increasing the spatial concentration of stops beyond a critical level tends to increase both the user and agency-related costs. This result demonstrates how excessive spatial coverage is counterproductive.

This model is reformulated in Badia et al. (2014) and applied to a radial route layout. Among the different outcomes, the authors showed that the radial layout is suitable for a centripetal demand pattern, in which the central area is the major attractor and generator of trips (like we assume in our work). The two studies show that a high-performance bus system (i.e., buses running on transit priority corridors) outperforms a rail rapid transit system for a wide range of demand density and coverage areas. However, since the former requires quite large streets, it appears unrealistic to imagine that such systems can entirely replace underground transit in the big cities' dense urban fabric.

In Chen et al. (2015), two different city-wide transit structures are compared, showing that the ring-radial layout is more favourable to transit (in terms of costs) than the grid design. However, the demand density is assumed to be spatially uniform over the entire urban area, which is not realistic. Note that none of the aforementioned studies in this subsection considers DR transportation.

Nourbakhsh and Ouyang (2012) proposed a transit network with no fixed routes: individual buses sweep back and forth through a tube-shaped predetermined area, where passengers are picked up or dropped off. Buses operate in a demand-responsive fashion in the respective "tube". The optimal structure parameters is obtained via a simple constrained nonlinear optimization problem. The authors showed that under low-to-moderate passenger demand the system incurs lower cost than other conventional counterparts such as the fixed-route transit system and the chartered taxi system. The system is however not suited for high demand. We instead observe that in a big conurbation, the demand can be high or low depending on the geographic sub-region and time-of-day considered and therefore it is not possible to just rely on DR transportation. For this reason, we instead keep FR operation at the core of transit and integrate DR to it.

The work discussed so far does not combine feeder in FMLM and trunk MRT, which is instead crucial for our *Adaptive Transit*. We discuss in the next section CA approaches for multimodal transit, which consider such a combination.

### 2.2.3 Continuous Approximation models for multi-modal transit

CA models have also been applied to multi-modal transit, with FMFL feeder and a trunk (or backbone), which corresponds in our design model to MRT. In these works, the feeder is either FR or DR. The novelty of our work is that we instead let our optimization decide between the two for each distance $x$ from the centre and for each time of day $t$, based on demand density.

Aldaihani et al. (2004) divided the study area in a grid, with a FR service along the lines of the grid and a DR service within each sub-region, consisting of a taxi service, serving one passenger at a time. We also divide the entire area in sub-regions, but we let our optimization choose between FRF and DRF therein. Moreover, our DRF is able to serve multiple passengers at a time. Sivakumaran et al. (2012) proposed a CA model to show the benefits of coordinating feeder services and MRT (trunk), but they only considered FRF.

Chen and Nie (2017) studied a grid and a radial network with fixed route transit lines integrated with DR transportation connecting passengers to the stops of the fixed lines. Optimal design is formulated as a mixed integer program. The results show that such a design outperforms the other two compared systems, one always using FR and the other always DR, under a wide range of scenario configurations. The main limit of Chen and Nie (2017) is that their DR service is designed to run over the entire urban area, everywhere with the same characteristics. e.g., with fixed headway. Our optimization problem decides instead where and when to deploy FRF or DRF (with the decision variable $F(x)$). Thanks to this optimization setup, we find that it is optimal to deploy FRF close to the city centre and DRF far from it, and how far depends on the time of day. Later, Luo and Nie (2019) compared six distinctive transit systems using the CA approach, most of them already studied in the previously cited works. A key finding is that the demand-responsive feeder services tilt the balance of trade-off considerably in the user's favour, at the transit agency's expense (which is consistent with our results). A recent work by Wu et al. (2020) compares fixed bus-based feeders with bike sharing-based feeders.

The systems proposed in past studies are far from being *adaptive*, i.e., they do not make an optimal decision between FR and DR modes, in each sub-region and period of day. Instead, such systems either use always one or the other. This paper instead shows that it is beneficial to make DR and FR co-exist in the same transit layout.

**2.3    Other approaches to integrated transit and demand-responsive transportation**

Salazar et al. (2020) proposed a network flow model of Integrated Autonomous Mobility on Demand (I-AMoD), where a ride sharing service provided via automated vehicles is integrated with transit. Ride sharing and transit are modelled together in a multi-layer graph. A static assignment problem is solved to calculate how the origin-destination matrix demand distributes onto the arcs of such a graph. Their goal is to find optimal pricing, while we aim to optimize the structure of the overall transportation system. Narayan et al. (2020), Leffler et al. (2021) and Bürstlein et Al. (2021) showed in simulation that DR feeders can improve access to line/schedule-based transit, with benefits in terms of quality of service and environmental impact. In simulation, Chouaki et al (2023) observed the increase of ridership of transit lines, when they are served by flexible feeders.

Chen et al (2020), Wen et al. (2018) and Shen et al. (2018) studied a feeder service provided by automated vehicles. The first resorted to mixed integer linear problem (MILP) to determine vehicle dispatch. The second included nested logit behavioural models and is based on simulation. The third, also based on simulation, assumed to keep only high-ridership feeder fixed lines serving the considered station and added autonomous vehicles to compensate for the other removed feeder lines; it then compared the resulting feeder system with the current one, entirely based on fixed lines. All the three papers do not explicitly build origin-to-destination passenger routes, but only deal with the part of the trip to/from the transit station. Ma et al. (2019) constructed instead multimodal user routes, which can include ride-sharing or walk (in the first and last mile) and fixed transit. Kim and Schonfeld (2014) studied via stochastic optimization a scenario with only one fixed transportation terminal, to which multiple First Mile/Last Mile (FMLM) sub-regions are connected via feeder buses, either fixed or flexible. Mahéo et al. (2019) proposed a system with few terminals (20 locations in their case); an unlimited feeder taxi fleet brings passengers to/from such terminals; the routes of fixed bus lines and multi-modal (taxi + bus) passenger journeys are calculated via integer programming.

Franco et al. (2020) generated demand for future DR services integrated with fixed transit, based on mobile data. Note that none of the studies mentioned in this subsection seeks to find the optimal transportation layout for an entire metropolitan area, which is instead our target. An and Lo (2015) solved the transit network design problem under demand uncertainty trough robust optimization for rapid transit and dial-a-ride services. However, the authors did not include passenger waiting times in the model and assumed that travel costs are proportional to distance (and not to travel time), which is unrealistic.

Pinto et al. (2020) proposed a model based on dynamic programming and simulation-based assignment. The decision variables they aim to calculate are two: the headways of bus lines and the fleet size *S* of a taxi-hailing service, in which each vehicle can have at most two riders on-board. The main difference of our work with respect to Pinto et al. (2020) is that we are interested in studying how the overall transit system can adapt spatially and temporally to the spatio-temporal demand variation, choosing in particular between FRF and DRF bus services. Instead, Pinto et al. (2020) let a mathematical program calculate a single value of *S*, without letting the agency decide in which regions and at which time of the day such *S* vehicles should be deployed. This suffers from the potential risk to just attract such vehicles in the city centres, where most of the demand is and where fixed transit is already efficient. This would play against our goal of employing DRF in low demand areas and during off-peak hours. We instead keep the choice of where and when to deploy DRF in the hand of the agency. Moreover, our DRF is able to serve many users (more than 2) at the same time and can act as a minibus or bus service.

The optimization problem of Steiner and Irnich (2020) aimed to "shorten" some bus lines, i.e., eliminate some stops at the beginning and the end of FR lines and replace them with a DR service. We believe that completely removing the FR service from the periphery of an urban area may worsen, rather than improving, mobility, overall. First, it would disadvantage suburban travellers even further. Second, it would require aggregating the demand close to the centre via a DR service, which may cause congestion and would suffer from limited capacity. We instead let the FR service to be deployed up to the extreme periphery of the urban area and adopt DR services as feeder, instead of as a replacement of FR lines. Finally, the authors did not show how the transit service should change configuration over the day to adapt to the time-varying demand pattern.

Fielbaum (2020) adopted an idealized parametric city model, which is not Continuous Approximation, but which has a comparable level of abstraction. His goal was to optimize a mixed transit structure, made of major fixed lines and automated feeder vehicles. While the decision variables concerning fixed lines are similar to ours, the feeder service is very different: they assume that each vehicle is assigned a specific single stop and operates back and forth from that stop to the closest fixed line stop. In order to start a trip, such a vehicle waits until its capacity is completely filled. This might be highly inefficient in areas where demand density is low, which are the ones that interest us the most. For this reason, we let our optimization model decide between a fixed-route or demand-responsive bus feeder. On the behavioural aspects, some authors as Anburuvel et al. (2022) have studied under which pricing and performance conditions flexible transportation would be preferred to fixed transportation in a developing country.

## 2.4 Positioning of our work

To the best of our knowledge, none of the previous work has tackled the problem of designing *Adaptive Transit*, i.e., to decide how to optimally vary spatially and temporally the layout of transit over an entire urban area, also deciding in which regions and in which time of day (peak / off-peak) FR or DR transportation must be deployed. A "variable" layout of this kind allows transit to better adapt to the demand, which is varying over time and space.

To this aim, we do not need to re-invent a model from scratch, but we build upon the previous work discussed in this section, in particular related to Continuous Approximation. We readapt it to our *Adaptive Transit* case as described in Section 4.

## 3. Transit Design Schemes

We focus on the transit system of wide urban and metropolitan areas and we assume it is formed by the following two components (as in the work reviewed in our Section 2.1):

- A Mass Rapid Transit (MRT).
- Possibly, a feeder service, provided by bus, to serve the First Mile and Last Mile (FMLM).

## 3.1 Central and suburban areas

The MRT network is modelled as a ring-radial structure, as in Badia et al. (2014) and Chen et al. (2015), which can be adopted to model several big cities around the world (e.g., Paris, Singapore, Moscow). As common in Continuous Approximation modelling, we assume the entire area is composed of two parts:

- A central area only served by MRT with double coverage provided by radial and ring rail lines.[3]
- A suburban area covered by MRT radial lines (and no ring lines).

---

[3] Note that, in case the central area is served by high frequency buses or bus rapid transit, our model would remain valid, with just modifications to the cost coefficients and capacity constraints. We just consider MRT in the central area to keep the model simple.

Feeder services can be deployed in the suburban area. Note that the sizes of the two areas above are not endogenous: they depend on the global decision variable $r$ (i.e., the radius of the central area - see Section 4.1), which is decided by the optimization procedure (see Section 4.7).

## 3.2 Transit schemes

We discuss three alternative transit schemes, which essentially differ in the way passengers can access MRT:

1. *MRT-only* scheme, in which the access to MRT stations can only take place by walking.
2. *MRT-FRF* scheme, which includes feeder bus lines with fixed routes to increase the accessibility of MRT stations in the suburban area. Such stations can be reached either by walking or using the fixed-route feeder bus, depending on the distance from the station.
3. *Adaptive Transit* scheme, in which the FMLM in the suburban area is still covered by a feeder service, but the feeder can switch between two modes, FRF and DRF, choosing optimally between one or the other based on the transportation demand density.

Figure 1 breaks down the components of the travel time of the passengers using transit. A passenger needs first to access transit. In case passengers use MRT, they can access it by either walking or using a feeder service (which can be a FRF or a DRF, depending on the scheme). A traveller is required to wait until a feeder vehicle arrives (*waiting time*) and spend some time in the vehicle (*in-vehicle travel-time for feeder*) to the MRT station, which results in an *access time* as depicted in Figure 1. Symmetrically, to reach the final destination from the egress MRT station, a traveller needs to walk or use another feeder service. Note that there is no walking time when the DRF is employed, as it assumed to be a door-to-door service.

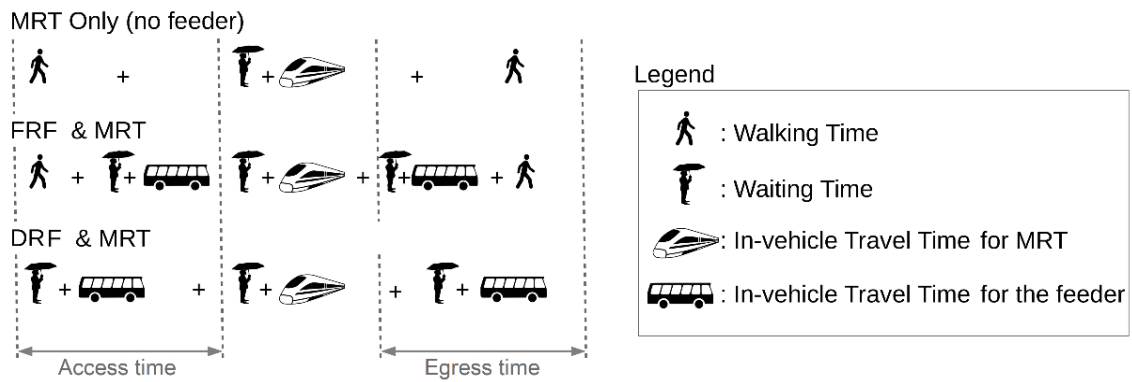

*Figure 1. Components of the access, egress and waiting time for the MRT-only scheme and when FRF or DRF services are provided.*

We clarify that in *Adaptive Transit* the choice of whether to deploy FRF or DRF is not made on-the-fly. On the contrary, we assume that, based on historical observation of the demand density, the authority would plan, for each area, the time periods when FRF or DRF will be operated. Such a plan would be revised only on a seasonal basis.

We use *MRT-only* as a baseline scheme. Its poor cost-efficiency shown later in the numerical results demonstrates that a feeder bus service in the suburban area is necessary. The *MRT-FRF* scheme is what it is basically currently deployed in most cities. The *Adaptive Transit* scheme is the design we propose for future generation transit.

## 4. Continuous Approximation Model

With the Continuous Approximation (CA) approach, an urban conurbation and a transit network are represented with a parametric model, consisting of:

- A set of decision variables, describing the layout of the transit network, i.e., the spacing between transit lines and stops, the value of the headway, etc.
- A set of input parameters, which are exogenous and describe the scenario, e.g., size of the urban conurbation, demand density.
- A set of constraints, which ensure basic properties, such as the conservation of flows and transit vehicle capacity constraints.
- A cost function, which we want to minimize; it includes a weighted sum of user-centric and agency-centric costs. It represents the performance of transit.

CA models allow understanding the impact of the different decision variables on the performance, in an approximated, concise and computationally efficient way. The results obtained via CA models should be interpreted as high-level trends, which can guide transit planning considerations. Therefore, we resort to CA modeling to understand the benefits of choosing between a FRF and a DRF service, in order to better adapt to the demand, over time and geographical areas, a concept that we call *Adaptive Transit*. We are not interested in exact results representative of a single specific city. For this reason, CA methodology perfectly fits our needs. Our formulation is mainly based on Chen et al. (2015); Daganzo (2010), both of which, however, do not integrate MRT and feeder. For this reason, we need to extend their models, as we will pinpoint in the following pages. Moreover, to model *Adaptive Transit*, it is crucial to let the model choose between FRF and DRF.

For this reason, we need to model both (based on Quadrifoglio and Li (2009)) and to add a binary decision variable to choose between the two.

One novelty with respect to the previous work on CA is that we introduce a notion of time-evolution. We need to do so, as we want to evaluate the capacity of our transit design to adapt to change of the intensity of the demand over the day. We therefore introduce a set $\mathcal{T}$ of time instants $t \in \mathcal{T}$, when the demand and supply of the transit system have given characteristics, and partition the entire day into non-overlapping time intervals starting in instants $t \in \mathcal{T}$, each of duration $\Delta t$ during which such characteristics remain constant. With slight abuse of notation, we will denote with $t$ a time instant and also the time interval starting at $t$.[4] The notation used in this section is summarized in Appendix A (Tables A1-A2).

## 4.1 Main decision variables

We study a circular metropolitan area of radius $R$ (exogenous input parameter). The transit layout, depicted in Figure 2, is organized as described in the previous section. It is described by 10 local decision variables and 4 global decision variables, determining the transit structure.

The local decision variables take a value for each value $x$ of distance (in km) from the centre. They are:

- The angle $\theta_r(x)$, in radiants, between radial MRT lines; based on that, we can also compute the corresponding linear spacing $S_r(x) = \theta_r(x) \cdot x$.
- The spacing $S_c(x)$ between ring MRT lines, defined for $x < r$, where $r$ is the radius of the central area.
- The spacing $s(x)$ between the MRT stations (hereinafter called just "stations") along a radial MRT line.
- The angle $\phi(x)$ between stations on a ring MRT line, defined for $x < r$.
- The headway $H(x)$ on ring and radial MRT lines.
- The headway $h(x)$ of the feeder service (only defined in the suburban area).
- The variable $F(x) \in \{\text{FRF,DRF,0}\}$ indicating whether at location $x$ a FRF, a DRF or no feeder service is deployed, respectively. We introduce the following indicator function $\mathbb{I}_j(x), j \in \{FRF, DRF, 0\}$, which is 1 for the $x$ where $F(x) = j$.
- The variable $d_{FRF}(x)$ (defined for $x > r$ and $F(x) = $ FRF) which is the spacing between FRF stops.

---

[4] Observe that, to keep the mathematical development treatable, we make the simplifying assumption that the aforementioned time intervals are independent of each other, in the sense that there is no propagation of passenger flows from one interval to the next. This is also equivalent to assuming that the flow starting in the current time interval and terminating in the next is compensated by the flow starting in the previous and terminating in the current.

- The variable $d_{0,DRF}(x)$ (defined for $x > r$ and $F(x) =$ DRF), which is the value determining the area close to the MRT station where the feeder service does not pick-up or drop-off passengers as we assume passengers would rather walk (see Figure 2).
- The number of strips $N_s(x)$, an integer variable defined for $x > r$, in which a FMLM sub-region is divided. Each strip is served by a feeder service. In the *MRT-only* scheme, $N_s(x) = 1$.

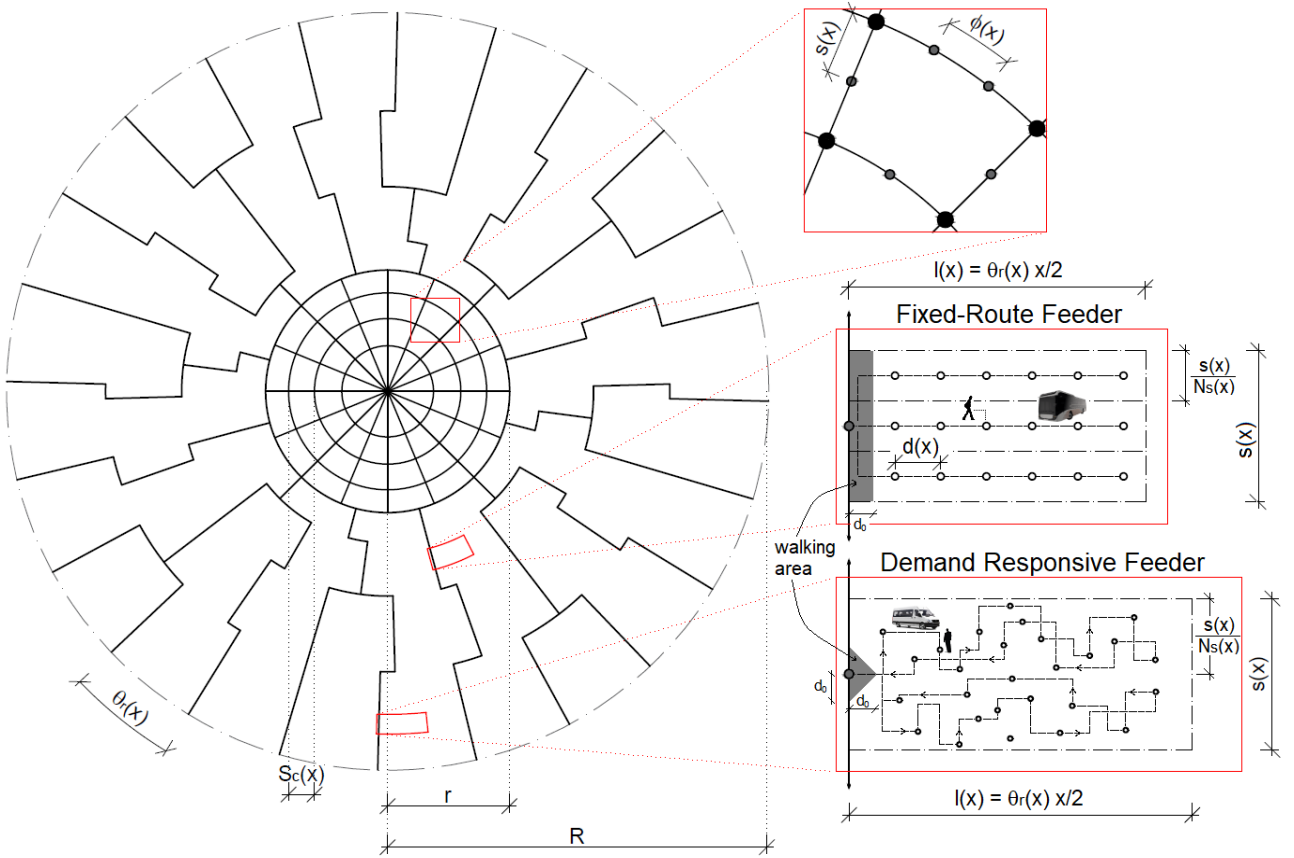

*Figure 2. Transit network layout (this takes inspiration from Chen et al. (2015) and Quadrifoglio and Li (2009)).*

The variation of such local variables along *x* can be seen as an approximation of what one could observe in reality. For instance, if radial lines bifurcate with the distance *x*, we can represent this by reducing the angular spacing $\theta_r(x)$ with *x*. Accordingly, the headway $H(x)$ would increase with *x* because less vehicles will travel along each line. Note that, to limit the number of parameters, we consider the same headway on ring and radial lines $H(x)$. This constraint is only active in the central area ($x < r$) and not in the suburbs, where no ring lines are present.

The global decsion variables are:

- The radius of the central area $r$.

- The angle $\phi_B$ between stations on the outermost ring MRT line, at $x = r$ (city centre's boundary).
- The headway $H_B$ of the MRT line at $x = r$. Observe that we need global variables for $\phi_B$ and $H_B$ because the outermost ring MRT line serves more trips than all other ring lines. Indeed, the outermost ring MRT line attracts the transfers of the travellers whose origin and destination are in the suburban area (see Figure 3b).[5]
- The maximum value $Q_0$ of the total radial flow of MRT vehicles, This maximum value occurs at $x = 0$.

Regarding the last decision variable, we define the radial flow $Q(x)$ as the number of MRT vehicles crossing an infinitesimal annulus of radius $x$ in both inward and outward direction.

$$Q(x) = \frac{2\pi}{\theta_r(x)} \cdot \frac{1}{H(x)} \quad (1)$$

Therefore, the two local decision variables $\theta_r(x)$ and $H(x)$ are interdependent. $Q_0$ is defined as $Q_0 \equiv Q(x = 0)$.

## 4.2 Assumptions and constraints

Along each radial MRT line, in the inward direction, we assume a train can depart from any $x \leq R$, but always terminate in the centre ($x = 0$). In the outward direction, a train always departs from the centre and can terminate at any $x \leq R$. This translates to the following constraint:

$$H(x_1) \leq H(x_2), \quad \forall\, x_1, x_2 \mid x_1 < x_2 < R \quad (2)$$

Also, we prevent the radial flow $Q(x)$ from increasing outward (it would mean that there were vehicles not passing through the city centre, contradicting our assumption) with the following constraint:

$$Q(x_1) \geq Q(x_2), \quad \forall\, x_1, x_2 \mid x_1 < x_2 < R \quad (3)$$

The following vehicle capacity constraint must also be respected:

$$O_j(x) < C_{pax,j}, \quad \forall x \quad (4)$$

---

[5] We can make a parallel with the variation of the electromagnetic field at the boundary between dissimilar media: in this case, we also need a mathematical development at the boundary different than at the other locations, since a discontinuity occurs there (Ellingson, 2021). The outermost MRT ring line acts as boundary between the city centre and the suburbs and a discontinuity, in broader terms, occur in this line, in the sense that it is used by much more passengers than all others ring lines, as it attracts all transfers of passengers whose origins and destinations are in the suburbs.

where $O_j(x)$ is the average vehicle occupancy at $x$ and $C_j$ is the vehicle capacity of mode $j \in$ {MRT, FRF, DRF} as computed in Equations (A.4)-(A.8) (Appendix A).

**4.3 Demand pattern and travel behaviour**

Metropolitan areas are characterized by a transition from a central zone to sprawled suburban areas. The former is characterized by dense urban fabric, high population density and presence of numerous "trip attractors" (job places, commercial activities, amenities, etc.). The latter, instead, are often shaped by low residential density and sparse transportation demand. We assume that the transit demand density is both temporally and spatially variable and follows the Clark's law (Clark, 1951), i.e., an exponential decline from the centre to the suburbs. We point out that the choice of a monocentric demand model is made to limit the number of parameters of the CA model. Other urban settings, e.g., polycentric and dispersed (Fielbaum et al., 2016), can be modelled to make the model more realistic, but would make our model more complex, which is outside the scope of this paper. The results may vary under different settings/assumptions, however our adoption of the Clark's law can be more applicable to general scenarios compared to previous (where the demand density is assumed constant (Daganzo, 2010) or with linear decrease (Badia, 2014)). Denoting with $x$ the distance (in km) from the city centre, the demand density (measured in trips per hour per km$^2$) is given by:

$$\rho(x,t) = \rho_0(t) \cdot e^{-\gamma x} \tag{5}$$

where $\rho_0(t)$ (pax/km$^2$ h) is the density of users in the centre and $\gamma$ (km$^{-1}$) is the slope (also called density gradient) with which that value decreases as we move away. By changing $\rho_0(t)$ over the time of the day, we can capture the temporal variation of the transportation demand. In our work we consider that $\rho_0(t)$ varies in a stepwise function, i.e., it remains constant within each time period $\in \mathcal{T}$. When we omit $t$, to simplify notation, it means we are focusing on a single time instant.

With such a model, the city centre, where economic activities are more concentrated, emerges as an attractor and generator of trips from/to the periphery (see Equation A.1). A passenger first accesses the closest transit station (either by walking or via a feeder bus), rides via MRT to the station closest to her destination and finally reaches (either by walking or via a feeder) her destination. Note that, within the MRT, a passenger could transfer from a radial to a ring line and vice versa. The sequence of such transfers obeys classic assumptions in literature (e.g., Badia et al., 2014; Chen et al., 2015) and is calculated to minimize the travelled distance via MRT. Transfers are developed in Appendix A.3 and summarized in Figure 3.

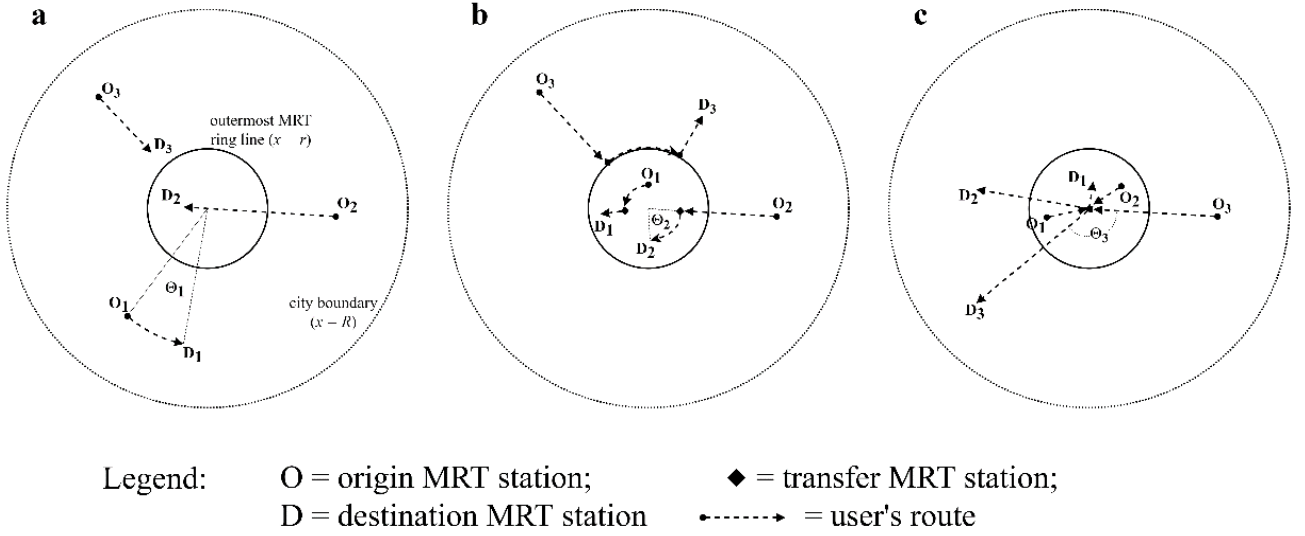

Legend:  O = origin MRT station;  ◆ = transfer MRT station;
D = destination MRT station  ┅┅→ = user's route

*Figure 3. User's route choice from the origin MRT station to the destination MRT station.*

In Figure 3a only those trips that do not require transfers between different MRT lines are represented. If we consider a pair, such as O2-D2 or O3-D3, they are both characterized by a small angle between O and D ($\Theta_1 \leq \theta_{r,min} = \min_{x \in [0,R]} \theta_r(x)$), passenger will just use one radial line only. If the angle is small and, additionally, O and D are at a similar distance *x* from the centre (like O1-D1), such a trip can be made without using MRT (see Appendix A.3). In Figure 3b we represent trips with angle $\Theta_2 \in ]\theta_{r,min}, 2$ radiants] between *O* and *D*. In this case, it is easy to see that, in order to minimize the travel distance, a user will travel via both radial and ring lines. Also, when both O and D lie in the periphery ($x > r$, e.g., $O_3$ and $D_3$ in Figure 3b), we have the only case that requires using the outermost ring line and implies two transfers. Finally, in Figure 3c the cases where the O-D pair has an angle $\Theta_3 > 2$ radians are shown: a user will travel by a radial line towards the city centre, where she will transfer to another radial line to reach her destination.

### 4.4 Feeder services

The suburban area is divided in FMLM sub-regions, each determined by the spacing $S_r(x) = \theta_r(x) \cdot x$ between the radial lines and the station spacing $s(x)$ along them, as in Figure 4. In case of *MRT-only* scheme, passengers can only walk inside each FMLM sub-region. In the other two schemes, instead, each FMLM sub-region is associated to a MRT station, and is further divided in a number $N_s(x)$ of strips served by a feeder bus service to/from that MRT station. Each FMLM sub-region, forming a ring sector, can be approximated into a rectangle with the following dimensions:

$$\text{length } l(x) = \frac{\theta_r(x)}{2} \cdot x; \quad \text{width } s(x); \quad \text{strip width } w(x) = s(x)/N_s(x) \tag{6}$$

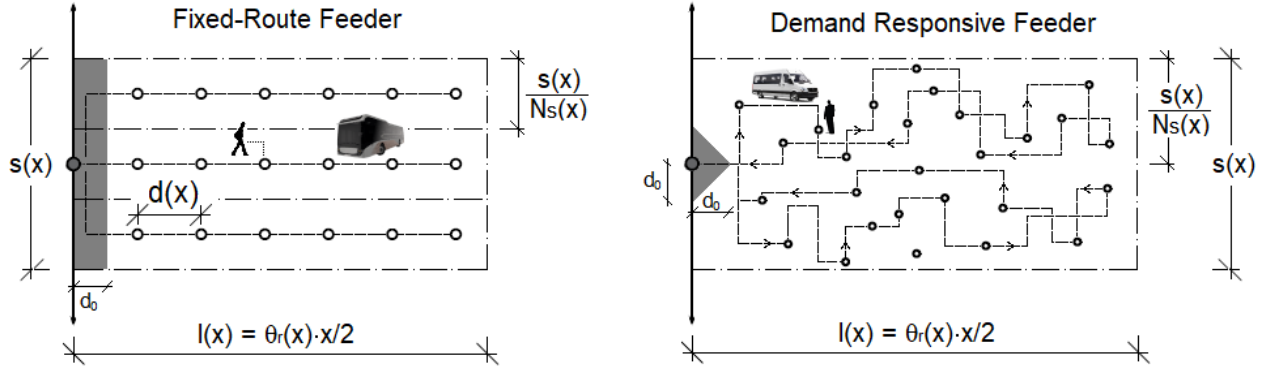

*Figure 4. FRF and DRF layouts.*

Note that the size of an FMLM sub-region depends on the MRT structure (the more the MRT lines and/or the smaller the station spacing, the smaller the FMLM sub-regions) and determines the total user demand to accommodate.

<u>Fixed-Route Feeder (FRF)</u>. The FRF is modelled as a straight route with spacing $d(x)$ between stops and vehicles moving back and forth between the MRT station and the furthest stop, as in Quadrifoglio and Li (2009), so that the length of a complete cycle is given by:

$$CL_{FRF}(x) = 2\left(l(x) + \Delta l(x) - \frac{1}{2}d(x)\right) \quad (7)$$

where $\Delta l(x)$ is the average extra vertical distance (see Figure 4) which the FRF has to travel due to the different position of the strips with respect to the MRT station they serve, that we approximate to $s(x)/4$ if $N_s(x) > 1$, and 0, otherwise.

We assume that travellers walk on a Cartesian grid, and thus all walking distances are Manhattan distances, which yields an average walking distance to reach the nearest bus stop equal to:

$$d_{FRF}^{walk}(x) = s(x)/4 + d(x)/4 \quad (8)$$

If the origin or the destination of a user is in a location close enough to the MRT station (less than threshold $d_{0,FRF} = d(x)/2$), she will prefer to directly walk to / from the MRT station. We call such locations "walking area", represented in grey in Figure 4. The fraction of locations in the walking areas is $p_{walk,FRF}(x) = d_{0,FRF}/l(x)$. And the time needed to complete a cycle can be calculated as follows:

$$C_{FRF}(x) = \frac{CL_{FRF}(x)}{v_{FRF}} + \tau_s\left(\frac{2l(x)}{d(x)} - 1\right) + \tau_p \cdot n(x) + \tau_T \quad (9)$$

where $v_{FRF}$ is the cruising speed of the bus, $\tau_T$ is the terminal dwell time, $\tau_s$ is the time lost per stop and $\tau_p$ is the time lost per passenger due to boarding/alighting operations. This time $\tau_p$ is multiplied by the average number of passengers per vehicle:

$$n(x) = 2\,\rho(x)\cdot w(x)\cdot l(x)\cdot h(x)\cdot\big(1 - p_{walk,FRF}(x)\big) \tag{10}$$

being $\rho(x)$ the demand density originating in any point at distance $x$ (in trips/km²h). We multiply by 2 to account for the passengers arriving at that point, under a symmetric demand assumption (arrivals and departures are the same at any point).

Demand-Responsive Feeder (DRF). The DRF model obeys the following assumptions. The DRF provides a door-to-door service, travelling along *x* and *y* direction grid, so passengers do not have to walk to any physical bus stop. Each request is processed in real-time via an insertion algorithm that aims at minimizing the in-vehicle travel time experienced by the passengers with a "no-rejection" policy, i.e., the DRF serves all assigned requests. If more passengers are assigned to a DRF vehicle (meaning more pick-ups/drop-offs), the vehicle needs to travel longer, due to longer detours on its service route (Oh et al. 2020b). Travellers close enough to the MRT station, i.e., at Manhattan distance less than $d_{0,DRF}(x)$ from the station (grey triangular area in Figure 4), directly walk to the station. Their fraction is given by the ratio between the surface of the walking area and the surface of the FMLM sub-region:

$$p_{walk,DRF}(x) = d_{0,DRF}^2(x)/(l(x)\cdot s(x)) \tag{11}$$

The computation of the cycle length ($CL_{DRF}$) and the cycle time ($C_{DRF}$) is based on the work of Quadrifoglio et al. (2006) and Quadrifoglio and Li (2009). The cycle length is estimated based on the expected number of passengers per vehicle $n(x)$, which is computed as in Equation 10, substituting $p_{walk,FRF}(x)$ with $p_{walk,DRF}(x)$. The cycle length is the sum of a horizontal component (the expected distance travelled from left to the right and vice-versa) and a vertical component (the expected deviations along the vertical direction to serve the passengers). The former is derived from Equation 9 of Quadrifoglio and Li (2009) and is equal to $2\cdot l(x)\cdot n(x)/(n(x)+1)$ assuming a uniform demand density within the FMLM sub-region. The latter is equal to $n(x)\cdot w(x)/3 + w(x)/2$ derives from Equations 3 and 4 of Quadrifoglio et al. (2006) (accounting for the fact that the vertical displacement occurs within a strip of width $w(x) = s(x)/N_s(x)$). The cycle length is thus:

$$CL_{DRF}(x) = 2\,l(x)\frac{n(x)}{n(x)+1} + n(x)\cdot\frac{w(x)}{3} + \frac{w(x)}{2} \tag{12}$$

The time needed to complete a cycle can be calculated as follows:

$$C_{DRF}(x) = \frac{CL_{DRF}(x)}{v_{DRF}} + (\tau_s + \tau_p) \cdot n(x) + \tau_T \tag{13}$$

The cycle time is the sum of the time travelling at cruise speed, the time lost per stop (assuming a single passenger served per stop) and the terminal dwell time.

### 4.5 Cost components

The main objective of the present work is to find the optimal transit structure able to integrate fixed and demand-responsive transportation. With this aim, we formulate a generalized cost function to be minimized as Badia et al. (2014) and Chen et al. (2015), which combines the disutility for users due to the travel time in its different components (Figure 1) and the costs incurred by the transit agency to provide the service (and the related externalities).

As regards the transit users, as usually done in CA work, we do not consider transit fares. This is a reasonable assumption when most users use monthly passes and since we do not consider a mode choice model in our work.

The quantities involved in the generalized cost are summarized in Table 1 and Table 2. They are all converted to monetary metrics via specific coefficients. They are all density functions over the distance from the centre. They are divided in two categories: (i) user-related costs represent the time spent and the discomfort suffered by passengers during their trip, both in the FMLM segments and in the MRT segment; (ii) agency-related costs, including capital and operational costs for operating feeder services in the FMLM and the MRT. The detailed computation of all cost components is in Appendices A.2–A.5.

Agency's and user's metrics are converted into cost density functions by means of a set of cost coefficients, in order to compute the total (per unit of time) as a linear combination of those metrics. We denote with $\mu_{L,MRT}$ or $\mu_{L,FMLM}$ (€/km-h), $\mu_{V,MRT}$ or $\mu_{V,FMLM}$ (€/veh-km) and $\mu_{M,MRT}$ or $\mu_{M,FMLM}$ (€/veh-h) the cost coefficients related to the agency metrics, for MRT and FMLM, respectively. The subscripts, have the same meaning of Table 1. Regarding the MRT, as in Flyvbjerg et al. (2013), we also consider costs specifically related to the stations through a coefficient $\mu_{ST}$ (€/Km-h). Similarly, cost coefficients $\mu_A$; $\mu_W$; $\mu_T$ are all equal to the Value of Time (VoT) (€/h) associated to walking, waiting and travelling on-board. For simplicity, we give them all the same unit, independent on whether they refer to travelling in a feeder or in the MRT. The cost components are detailed in Table 1 and Table 2, and discussed in next sections.

## 4.6 Optimization problem

To express the cost objective in a concise way, we use $i$ to indicate the type of cost component, $i \in \{L, ST, V, M, A, W, T\}$ (see Table 1), emphasizing that FRF and DRF are mutually exclusive feeder services. As in Chen et al. (2015), we distinguish:

- Local densities $Y_i(x)$, referred to the MRT, and $y_i(x)$, referred to the FMLM (either FRF or DRF), which vary with the distance from the centre $x$.
- Global components $F_i$ which are instead only related to the outermost MRT ring line, i.e., at $x = r$.

*Table 1. Overview of agency-related and user-related local densities.*

|  |  | FMLM |  | MRT |
|---|---|---|---|---|
| User-related costs | $y_A(x)$ | Cost due to walking to/from the feeder bus stop | $Y_A(x)$ | Cost due to walking to/from the MRT station |
|  | $y_W(x)$ | Cost due to the time to wait for the feeder service | $Y_W(x)$ | Cost due to the time to wait for the MRT |
|  | $y_T(x)$ | Cost due to the time spent into feeder vehicles, including boarding, riding, dwell, and alighting time. | $Y_T(x)$ | Cost due to the time spent into MRT, including boarding, riding, dwell, and alighting time |
| Agency capital costs | $y_L(x)$ | Cost for the infrastructure in FMLM, i.e., construction and maintenance | $Y_L(x)$ | Cost for the infrastructure of the MRT, i.e., construction and maintenance |
|  | $y_M(x)$ | Cost due to the feeder fleet, i.e., vehicles and crew cost | $Y_M(x)$ | Cost due to the MRT fleet, i.e., vehicles and crew cost. |
|  |  |  | $Y_{ST}(x)$ | Cost due to the MRT station density. |
| Agency operation costs | $y_V(x)$ | Cost due to vehicle-distance travelled by feeder vehicles. | $Y_V(x)$ | Cost due to vehicle-distance travelled by MRT |

*Table 2. Overview of agency-related and user-related global components*

|  |  | MRT |
|---|---|---|
| User-related costs | $F_A$ | Transfer cost, due to users' changing MRT lines at the outermost ring line (Figure 3b) |
|  | $F_W$ | Cost due to the time to wait for the MRT at the outermost ring line |
|  | $F_T$ | Cost due to the time spent into MRT along the outermost ring line |
| Agency capital costs | $F_M$ | Cost due to MRT fleet and crew cost on the outermost ring line |
| Agency operation costs | $F_V$ | Cost due to the vehicle-distance travelled on the outermost ring line |

We now formulate the optimization problem that we aim to minimize. We separate the decision variables in two sets:

- Global decision variables: $G = \{r; Q_0; \phi_B; H_B\}$,
- Local decision variables (functions of $x$): $D(x) = \{\theta_r(x); S_c(x); s(x); \phi(x); H(x); h(x); d_j(x); N_s(x); F(x)\}$.

As in Chen et al. (2015), we constrain the station spacing on the outermost MRT ring line to be the same as the value of the corresponding local variable at $x = r$:

$$\phi_B = \phi(r) \tag{14}$$

For any values of sets $G$ and functional $D = \{D(x)\}_{x \in [0,R]}$ (denoting all values $D(x)| x \in [0, R]$) at any time interval $t \in \mathcal{T}$, we denote the hourly cost of component $Z_i$ as:

$$Z_i(D, G, t) = \mu_{i,MRT} \cdot \left( F_i(r, \phi_B, H_B, D, t) + \int_0^R Y_i(D, r, t, x) \, dx \right) + \mu_{i,FMLM} \cdot \int_0^R y_i(D, r, t, x) \, dx \tag{15}$$

Note that all cost components $y_i(x)$ and $Y_i(x)$ depend not only on $x$ but also on the demand density $\rho_0(x)$, as it will be clarified in Appendix A, where such cost components are calculated.

As regards agency-related costs, they are composed of capital costs, which are not dependent on the time of day, and operation costs, which instead vary with $t$. The total cost is:

$$Z(D, G, t) = Z_{user}(D, G, t) + Z_{cap}(D, G) + Z_{op}(D, G, t) = \sum_{i \in \{A,W,T\}} Z_i(D, G, t) + \sum_{i \in \{L,M\}} Z_i(D, G) + \sum_{i \in \{V\}} Z_i(D, G, t) \tag{16}$$

For each of the three schemes defined in Section 3 (*MRT-Only*, *MRT-FRF*, *Adaptive Transit*), we first dimension the system in the peak hours. To do so, let $t^{peak}$ be the time interval in which the demand density is the highest, i.e., $t^{peak} = \arg\max_t \rho_0(t)$. We want to find the optimal set of global decision variables $G^{peak}$ and local decision variables $D^{peak}$ (for every value of $x$, i.e., $D^{peak} = \{D^{peak}(x)\}_{x \in [0,R]}$) by solving the following optimization problem:

$$\{G^{peak}, D^{peak}\} = \arg\min_{G,D} \left\{ \begin{array}{c} Z(D, G, t^{peak}) \\ \text{subject to Equations 2;3;4} \end{array} \right\} \tag{17}$$

This optimization allows us to determine optimal fleet size and transit infrastructure. Note that, once fleet size and transit infrastructure are fixed to satisfy the mobility needs in peak hours, they do not change over the day, and the corresponding cost must be supported by the agency, even if there are periods of the day in which they are not fully used. We thus fix the capital cost as:

$$Z_{cap} = \sum_{i \in \{L,M\}} Z_i(D^{peak}, G^{peak}) \tag{18}$$

We then optimize the system in each time slot independently, only minimizing the operational cost:

$$G^*(t), D^*(t|Q_0, r) = \arg\min_{G,D} \begin{Bmatrix} Z_{op}(D, G, t) + Z_{user}(D, G, t) \\ \text{subject to Equations 2;3;4;20(a-g)} \\ \text{with max flow } Q_0 \text{ and central radius } r \end{Bmatrix} \quad (19)$$

where we introduced the following constraints, valid when $t \neq t^{peak}$:

    a. $r = r^{peak}$

    b. $\theta_r(x) = \theta_r^{peak}(x) \quad \forall x$

    c. $S_c(x) = S_c^{peak}(x) \quad \forall x < r$

    d. $s(x) = s^{peak}(x) \quad \forall x$   (*MRT-only* or *MRT-FRF* scheme)

    e. $s(x) = s^{peak}(x) \quad \forall x < r, \quad s(x) \geq s^{peak}(x) \quad \forall x > r$   (*Adaptive* scheme)

    f. $d_j(x) = d_j^{peak}(x) \quad \forall x > r, j \in \{FRF, DRF\}$   (*MRT-FRF* scheme)

    g. $d_j(x) \geq d_j^{peak}(x) \quad \forall x > r, j \in \{FRF, DRF\}$   (*Adaptive* scheme)   (20)

Note that Equations 20(a-f) represent the fact that the infrastructure remains, all over the day, the same as the one decided via Equation 17. The total cost, over the entire day, is:

$$Z^{24h} = \sum_{t \in \mathcal{T}} \Delta t \cdot \left( Z_{cap} + Z_{op}(D^*(t), G^*(t), t) + Z_{user}(D^*(t), G^*(t), t) \right) \quad (21)$$

The optimization procedure is done separately for *MRT-only*, *MRT-FRF* and *Adaptive Transit* schemes.

### 4.7 Optimization procedure

We now describe the optimization procedure executed for each time instant. The optimization of Equation 17 and Equation 19 is non-convex, so we resort to bi-level optimization to solve it. Let us fix any time instant $t$. The lower level subproblem consists, given any value of global variables $r$, $Q_0$, to solve the following local optimization problem, for all $x \in [0, R]$.

$$D^*(x, t|Q_0, r) = \arg\min_{D(x)} \begin{Bmatrix} \sum_i \mu_i \cdot (Y_i(D(x), r, t, x) + y_i(D(x), r, t, x)) \\ \text{subject to Equations 2;3;4;20(a-g)} \end{Bmatrix} \quad (22)$$

We use an interior-point algorithm to solve the problem above. The higher level subproblem (global optimization) is to determine the set of global variables $G(t)$ that minimizes the total cost $Z$, which is the sum of global and local costs (Equation 15). The following iterative procedure is implemented:

1. Initialize $r$ (to a sufficiently small value, e.g., 3 Km,[6] and repeat the overall optimization procedure (described in step 3) by increasing $r$ until the total cost found is higher than the average value from the previous 3 iterations.[7] When this occurs, set $r^*$ equal to the value of the third to last iteration.
2. For any value of $r$, initialize $Q_0$ (e.g., $Q_0 = 100$ vehicles/h) and, similarly to the previous point, increase $Q_0$, until the total cost found is higher than the average value from the previous 3 iterations. When this occurs, set $Q_0 = Q_0^*$ the value of the third to last iteration.
3. For any pair of $(r, Q_0)$, run the lower-level optimization procedure, composed by the following steps:
   a. Run the local optimization (Equation 22) in order to find $D^*(x,t|Q_0,r)$ for all $x \in [0, R]$ (we discretize this interval with a step $\Delta x = 0.5$ Km).
   b. Find $H_B^* = \arg\min_{H_B}\{\sum_i \mu_{i,MRT} \cdot F_i(r, \phi_B, H_B, D^*, t)$ subject to Equation 4$\}$ . This problem is simple to solve for a solver since it has only one decision variable.[8]
   c. Compute the total cost as in Equation 16.

Observe that we first run the optimization procedure explained above in the peak hours. We then repeat the optimization procedure for all the other time instants, with the additional constraints of Equation 20(a-g), which are needed to keep infrastructure length and fleet size fixed to the peak hour values.

## 5. Numerical results

We now compare the performance of the three transit schemes (Section 3) in a scenario representing a large urban conurbation, during peak and off-peak hours. Note that we also present the benefits of *Adaptive Transit* in a separate small-scale simulation scenario (Appendix B).

### 5.1  Scenario parameters

The complete list of the parameters describing our scenario is reported in Table 3. We consider a circular area of radius $R = 25$ km, which corresponds to the size of large metropolitan areas, e.g., the *Greater Paris* region. The value of the demand density $\rho_0(t)$ is estimated from the travel demand data of the regional Household Travel Survey "EGT 2010".[9] We approximately fit Equation 5 on residential density data on Paris region,[10] which results in an average demand density $\bar{\rho}_0 = 640$ trips/(km² h) (this is the sum of trips departing from and arriving at each km²) and a slope of $\gamma = 0.12$ km⁻¹. The shape of the transit demand as a function of the distance from the city centre is represented in Figure 5a.

---

[6] Observe that at first we run the entire optimization procedure for $t = t^{peak}$. During this optimization, we explore different values of $r$, by trying different initialization values. Then, for all the other $t \neq t^{peak}$, we instead always fix $r = r^{peak}$.
[7] We consider the previous three iterations instead of just the last one to make our optimization procedure more robust.
[8] Note that $\phi_B = \phi(x = r)$ (Equation 14).

*Table 3. Parameters of the base scenario*

| Parameter | Value | Reference |
| --- | --- | --- |
| $R$ | 25 km | - |
| $\rho_0(t^{peak})$ | 1600 trips/(km²h) | - |
| $\rho_0(t^{op})$ | 480 trips/(km²h) | - |
| $\rho_0(t^{lp})$ | 256 trips/(km²h) | - |
| $\gamma$ | 0.12 km⁻¹ | - |
| $v_w$ | 4.5 km/h | *Google Maps* |
| $v_{MRT}$ | 60 km/h | Daganzo (2010) |
| $v_{FRF}, v_{DRF}$ | 25 km/h | Daganzo (2010) |
| $C_{MRT}$ | 1200 | - |
| $C_{FRF}, C_{DRF}$ | 80 | - |
| $\tau_{s,MRT}$ | 45 s | Daganzo (2010) |
| $\tau_{s,FRF}, \tau_{s,DRF}$ | 30 s | Daganzo (2010) |
| $\tau_p$ | 2 s | - |
| $\mu_A, \mu_W, \mu_T$ | 15, 22.5, 30 €/h | Meunier and Quinet (2015) |
| $\Delta A$ | 2 min | - |
| $\mu_{L,MRT}$ | 600 €/km h ($x < r$); 300 €/km h ($x > r$) | Flyvbjerg et al. (2013) |
| $\mu_{ST,MRT}$ | 300 €/(station h) (for $x < r$) 100 €/(station h) (for $x > r$) | Flyvbjerg et al. (2013) |
| $\mu_{L;FRF}, \mu_{L,DRF}$ | 10 €/(km h) | CERTU (2011) |
| $\mu_{V,MRT}$ | 6 €/(veh km) | Daganzo (2010) |
| $\mu_{V,FRF}, \mu_{V,DRF}$ | 0.5 €/(veh km) | Cats and Glück (2019) |
| $\mu_{M,MRT}$ | 120 €/(veh h) | Daganzo (2010) |
| $\mu_{M;FRF}, \mu_{M,DRF}$ | 50 €/(veh h) | Cats and Glück (2019) |

---

[9] The travel demand of *Greater Paris* consists of 8.3 million PT trips, on average per working day. Since a complete trip needs on average one transfer, we obtain a daily demand of 4.15 million trips made by PT per day. Considering 18 hours of PT operation, we obtain an average demand of 230·10³ trips/h.
[10] https://www.insee.fr/fr/statistiques

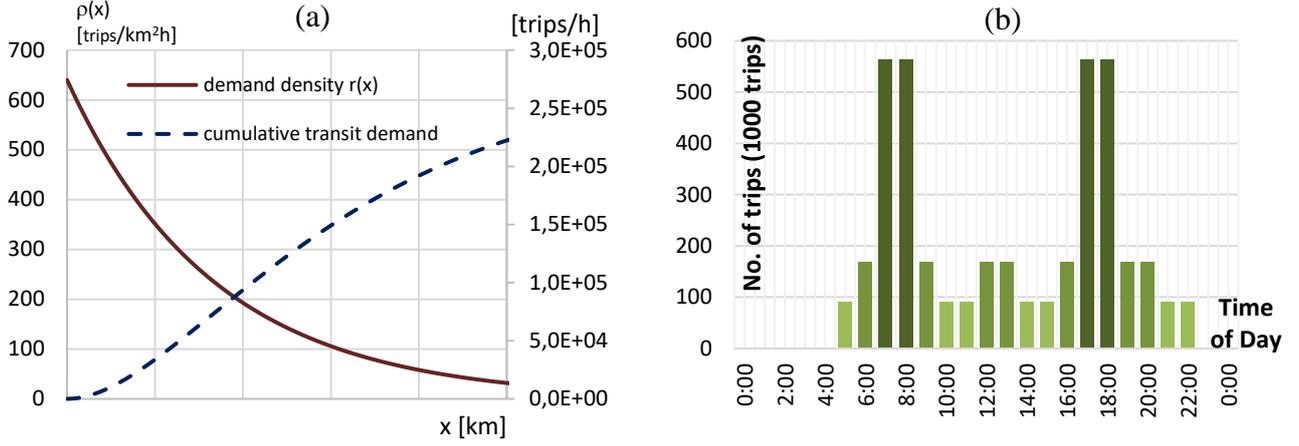

*Figure 5. (a) Demand density and cumulative transit demand as functions of the distance from the city centre; (b) Transit demand fluctuation during the day.*

For the sake of simplicity, we divided the day into time intervals $\Delta t = 1$ hour, and assume that $\rho_0(t)$ only takes three values during the day: a peak value $\rho_0(t^{peak}) = 2.5 \cdot \bar{\rho}_0 = 1600$ trips/km²h; an off-peak value $\rho_0(t^{op}) = 0.75 \cdot \bar{\rho}_0 = 480$ trips/km²h; and a low-peak value $\rho_0(t^{lp}) = 0.4 \cdot \bar{\rho}_0 = 256$ trips/km²h. Using this scheme, which is represented in Figure 5b, we obtain a ratio between $\rho_0(t^{op})$ and $\rho_0(t^{peak})$ of 3/10, as suggested by Jara-Díaz et al. (2017).

Referring to Table 3 we assume the cost coefficients of the FRF and the DRF are equal: this is because we assume that the switching between the two feeder services, in the *Adaptive Transit* scheme, occurs maintaining the same vehicles. Also, we set the cost coefficient $\mu_{L,MRT}$ in the suburban area is half of the one in the central area, since MRT lines do not require extensive tunnelling (which greatly impacts the costs) outside the city centre. For the same reason, the cost coefficient $\mu_{ST,MRT}$ in the central area is 3 times larger than outside (see Flyvbjerg et al., 2013).

Finally, to obtain the capital cost coefficients of Table 3, we use a straight-line amortization assuming 20 years (of 365 days) of useful life for MRT, 12 years for FRF and DRF buses,[11] considering 18 operating hours per day.

## 5.2 Performance of *adaptive* design scheme

We now compare the overall cost $Z^{24h}$ obtained with the three transit schemes, the difference in their optimal structure and the impact on user QoS, to show the superiority of our proposed *Adaptive Transit* over conventional transit design. The results are obtained, for each transit scheme, by applying the optimization procedure (Section 4.7) on the respective CA model. Such procedure searches for the optimal structure, i.e., the set of values of the decision variables that minimize the total cost (Equation 21).

---

[11] New York Metropolitan Transportation Authority: https://new.mta.info/document/11976

We implemented our procedure in MATLAB and release the code as open source.[12] It is computationally efficient, as it takes about 10-20 minutes in an ordinary laptop for optimizing each transit scheme. As a comparison, observe that one single agent-based simulation in Narayan et al. (2020), took 45h on a super-computer. Obviously, we expect that the accuracy of their results is much stronger than ours. However, such huge computation times are not suitable to analyse high-level managerial insights and when a fast way to experiment with different system parameters is needed.

Figure 6 (left) shows the most relevant differences between the performances of conventional transit and the *Adaptive Transit* scheme. Let us partition the time periods into disjoint sets $\mathcal{T} = \mathcal{T}^{peak} \cup \mathcal{T}^{op} \cup \mathcal{T}^{lp}$. We represent the capital cost $Z_{cap} = (Z_L + Z_M)$ (defined in Equation 18), the operational cost $Z_{op} = Z_V$ and the user-related cost components $Z_{user} = Z_A + Z_W + Z_T$. We recall that the total cost $Z^{24h}$ over the entire day is computed via Equation 21. It is clear that *Adaptive Transit* greatly reduces user-related cost in all periods of day. Particularly, more reduction is observed outside the peak hours, where classic fixed transportation shows more evidently its limitations. The reduction in user-related cost is mostly achieved thanks to a remarkable reduction of access (walking) time $Z_A$. Indeed, where and when the demand density is low (suburbs, off-peak), *MRT-only* scheme provides only few stops, to prevent the operational cost to explode, thus requiring users to walk long distances. A similar (although less pronounced) problem occurs in Off- and Low-peak with *MRT-FRF scheme*. Instead, *Adaptive Transit* does not bring such an issue, as when and where FRF becomes too disadvantageous for users, it deploys DRF, which picks up and drops off users at their locations. Observe that this improvement for users (-21.8% and -8.7% with respect to the *MRT-only* and the *MRT-FRF* schemes, respectively) requires higher agency-related costs (+ 1.1% with respect to the *MRT-FRF* scheme), both capital cost (fleet $Z_M$ and infrastructure $Z_L$) and operational (vehicle-km cost $Z_V$), since the agency needs to deploy more feeder vehicles.

However, such an increase of agency-related cost may be acceptable, as the overall cost of *Adaptive transit* outperforms the other schemes: Figure 6 shows that *Adaptive transit* reduces the overall total cost (-19.7% and -3.6% with respect to the *MRT-only* and the *MRT-FRF* scheme, respectively), in particular during the off-peak hours (7.2% of improvement compared to the *MRT-FRF* scheme).

---

[12] https://github.com/giovanni-cal/future-transit

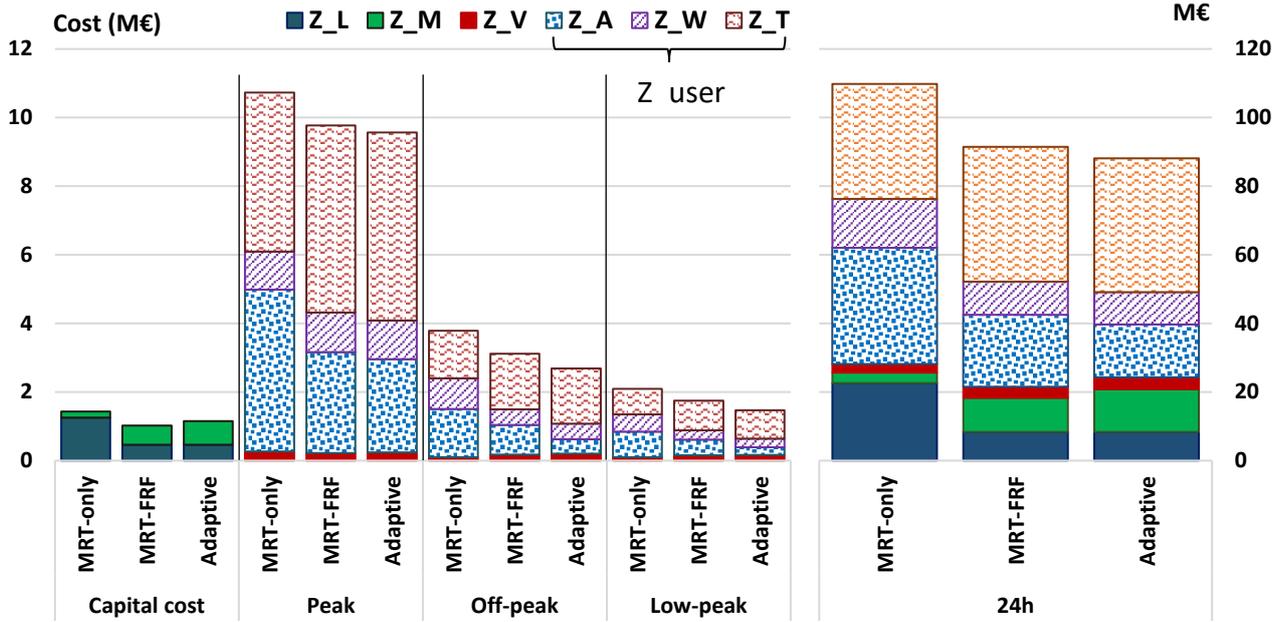

*Figure 6. Cost of MRT-Only, MRT-FRF and Adaptive scheme across the day (for result interpretation; $Z_L, Z_M, Z_V$ are the costs due to infrastructure, fleet (and crew), and operation, respectively; $Z_A, Z_W, Z_T$ are the costs due to walking, waiting and in-vehicle travel time, respectively).*

## 5.3 Spatial distribution of cost

In Figure 7, we divide the study area in 3 different zones: the first zone for x ≤ 6 km, the second zone for 6 km < x ≤ 15 km, and the third for x > 15 km. Such zones have comparable size to *Paris city*, *Petite Couronne* and *Grande Couronne*, respectively. With this figure we aim at evaluating how the user-related cost (in terms of travel time) distributes among the three zones in the different design schemes. We represent in each zone the average time (in minutes) suffered by the users of the transit system to access the MRT station.

We observe that the time needed to access the MRT explodes far the from city centre, in particular during off-peak hours, in the *MRT-only* and *MRT-FRF* scheme. In the first case, the discomfort for passengers far from the centre is exacerbated by high waiting times.

*Adaptive Transit*, instead, compensates the increase in waiting time by guaranteeing a fast connection to MRT stations far from the centre, i.e., x > 15 km (-9.4%, -32.1%, and -34.1% of access time with respect to the MRT-FRF scheme during peak, off-peak and low-peak hours, respectively). Therefore, it prevents access to MRT from degrading in suburban areas. Our *adaptive* design alleviates the cost suffered from users in the suburbs much more than conventional designs, by shifting the agency-related costs toward the outskirt (see the next subsection). Observe instead that conventional design suffers from a bias, favouring the city centre, in cost distribution: the agency invests more in the city centre, so that user-related cost is minimized there, at the detrimental of suburban population. In order words, conventional designs inherently suffer from high inequality. Such an inequality (Badeanlou et al., 2022) is alleviated with *Adaptive Transit*, which improves user-cost in the suburbs, without excessively degrading performance in the centre.

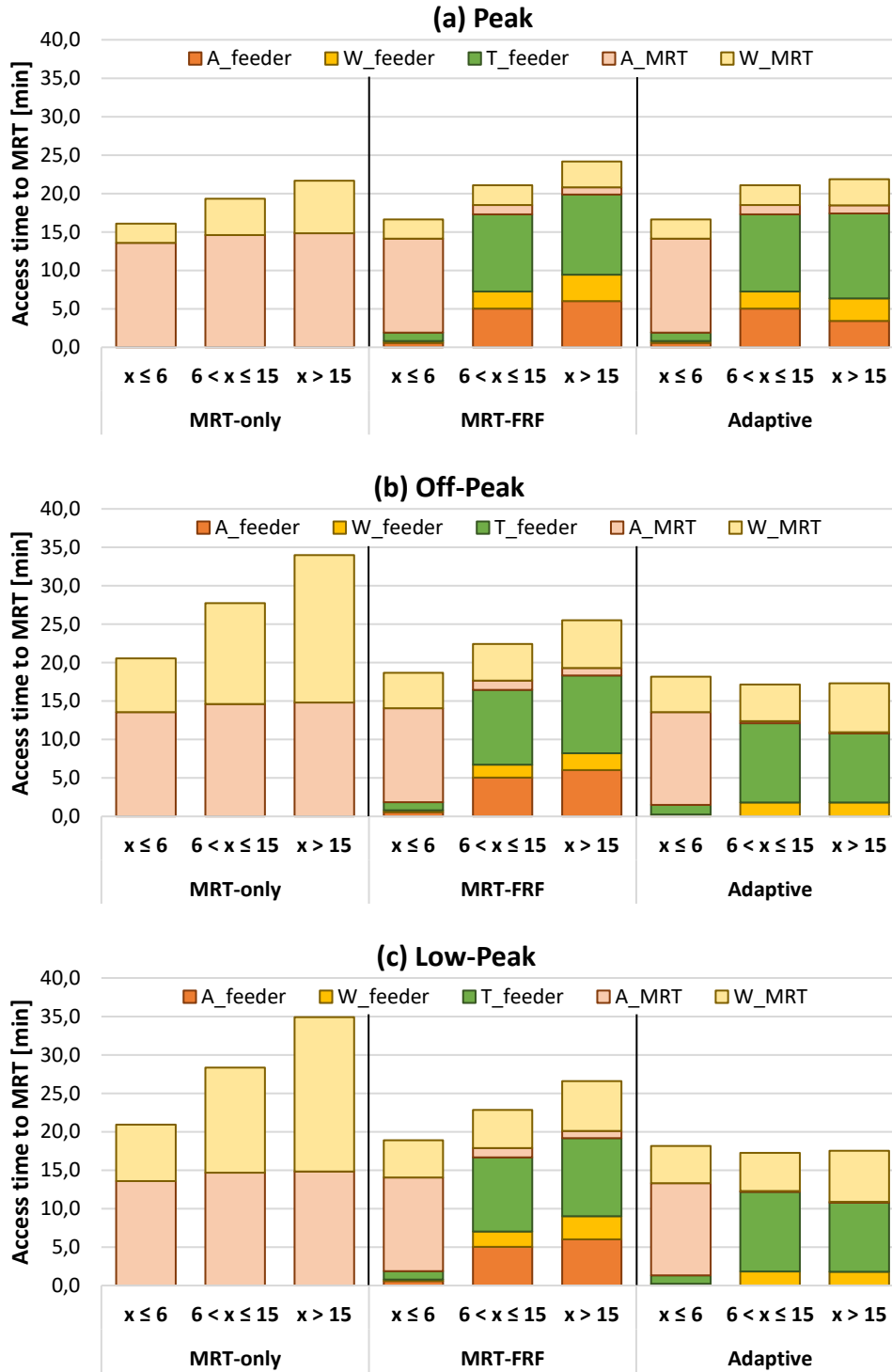

*Figure 7. Components of the total access time to MRT stations in 3 zones of the study area.*
*This time is made up by five components, i.e., the walking time $A_{feeder}$ to access/egress the feeder stop (if any), the waiting time $W_{feeder}$ for the feeder service (if any), the in-vehicle travel time $T_{feeder}$ on the feeder (if any), the walking time $A_{MRT}$ to access/egress the MRT station and the waiting time $W_{MRT}$ at the MRT station.*

## 5.4 Spatial adaptivity of *Adaptive Transit*

Figure 8 shows the optimal structure of the three transit schemes with reference to the peak and the off-peak hours, explaining the results previously discussed. Observe that the three schemes slightly differ in the central area ($x < r$), since in any case only MRT is deployed there.

We observe that the optimal value of $r$ for the *MRT-only* scheme (Figure 8a-b) is considerably higher than the schemes adopting feeder services in the FMLM. In fact, the outermost MRT ring line should have a 9 km radius vs. a 5.5 km radius of both *MRT-FRF* (Figure 8c-d) and *Adaptive Transit* (Figure 8e-f). This ensures a wider double-coverage (ring and radial lines) area for *MRT-only*, but it results, as already shown, in higher infrastructure-related cost.

The difference between the three schemes is more pronounced in the suburban area. In Figure 8c one can see that deploying FRF services allows the transit agency to save on infrastructure cost (by increasing the spacing $S_r(x)$ between radial MRT lines) and also to halve the headway $H(x)$ of the MRT with respect to the *MRT-only* scheme. Moreover, the distance between MRT stations is higher in *MRT-FRF* than in *MRT-only*, because users can exploit the feeder service instead of walking. As one can note from Figure 8d, during off-peak hours MRT headway $H(x)$ is almost twice higher than during peak hours (Figure 8c), while feeder headway $h(x)$ is reduced. Such an improvement in the frequency of the feeder service compensates the higher waiting time for the MRT during off-peak and low-peak periods.

Finally, Figure 8e and Figure 8f show the decision variables derived through the optimization process for peak and off-peak hours for the *Adaptive Transit* scheme. During peak hours, *Adaptive Transit* deploys FRF in the close suburbs, where the demand is sufficiently high, and relegates DRF only to the further periphery ($x > 21\ km$), where the feeder service sub-regions are slightly smaller compared to $x < 21\ km$ (see the values of $S_r(x)$ and $s(x)$ in Figure 8(e)). Moreover, the DRF requires a lower headway $h(x)$ to better accommodate the demand.

Note that lower values of feeder headway $h(x)$ do not imply a higher fleet size, since the number of operating vehicles depends also on the cycle length $CL_{DRF}(x)$ (based on the number of requests to serve) and the size of the FMLM sub-regions.

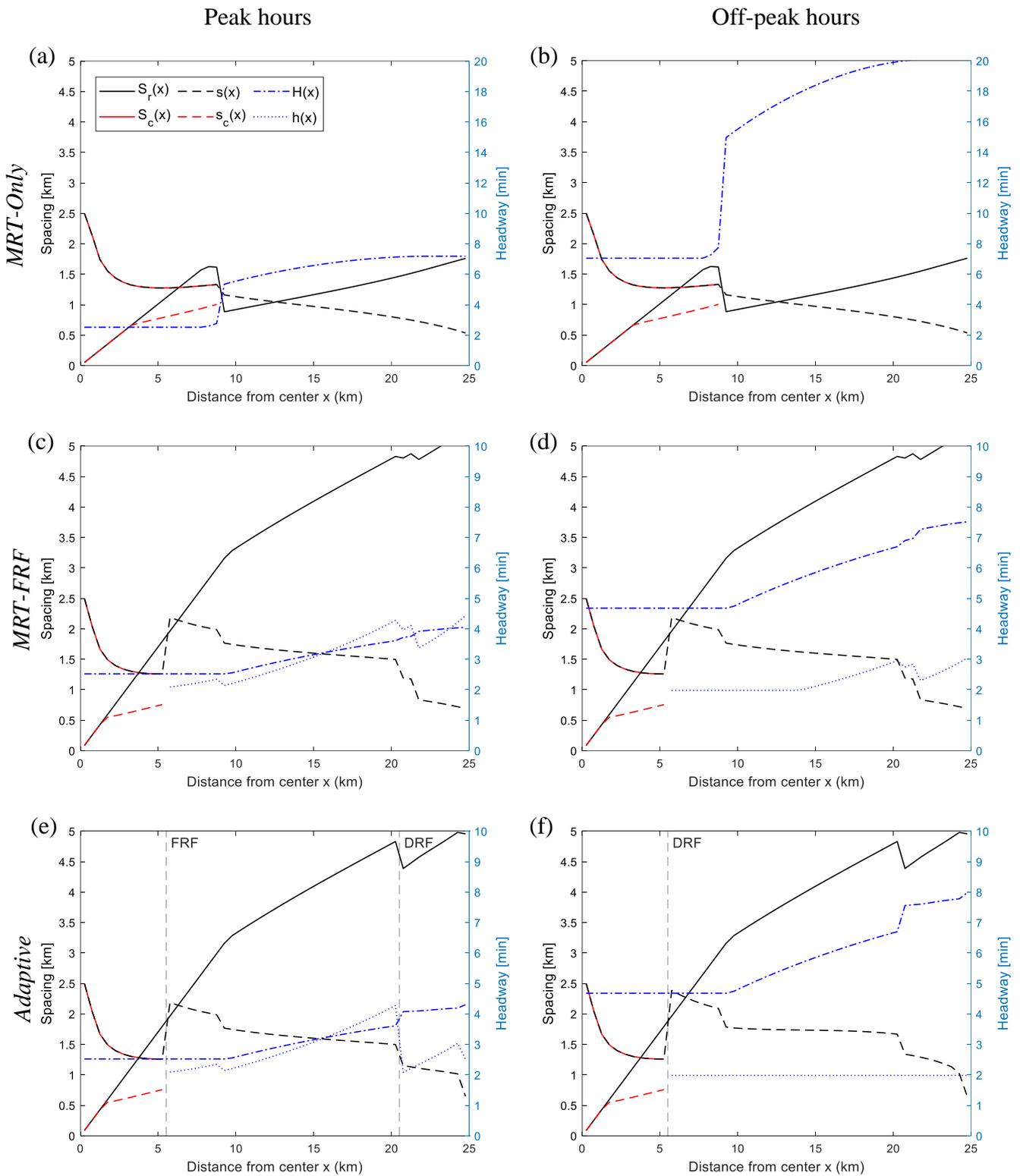

*Figure 8. Optimal decision variables for the transit schemes: MRT-only during (a) peak and (b) off-peak hours; MRT-FRF during (c) peak and (d) off-peak hours; Adaptive during (e) peak and (f) off-peak hours (for result interpretation; $S_r(x)$ = spacing between radial MRT lines; $S_c(x)$ = spacing between ring MRT lines; $s(x)$ = spacing between MRT stations along a radial line; $s_c(x)$ = spacing between MRT stations on a ring line; $H(x)$ = headway on MRT lines; $h(x)$ = headway of the feeder services).*

Observe that, in all the schemes, MRT offer is richer where the demand density is high, i.e., the headway $H(x)$ and the radial line spacing $S_r(x)$ are smaller in the suburbs closer to the centre than in the furthest suburbs. This is also what we observe in real cities. We also observe that the further we go from the centre, the more stop spacing $s(x)$ decreases. This trend is more evident in the peak hours, due to the fact that stop spacing strongly affects the commercial speed of MRT and, consequently, the total travel time of passengers. Since the demand density is higher close to the city centre, it is more convenient to have a faster service (with less stops) in those areas.

It is important to remark how the structure of our *adaptive* structure changes over the time and space. It is worth highlighting how, for $x > r$, due to the low demand density, *Adaptive Transit* deploys DRF outside the peak. Moreover, the spacing between MRT stations increases going from peak to off- and low-peak hours. We pinpoint that such variation should not be interpreted as an unreal modification of the infrastructure, but as an operational strategy consisting of "skipping" some stops in certain periods of the day, i.e., introducing an "express" service, serving only a subset of all the possible MRT stations, in order to offer faster connections. This shows that *Adaptive Transit* is able to vary the deployment of feeder services spatially, to adapt to the geographic demand gradient, and temporally, to adapt to the time-evolution of the demand.

## 5.5 Effects of varying urban area size and value of time

We now evaluate the overall cost under difference exogenous scenario parameters. In particular, we study the impact of the radius $R$ of the city and the VoT for passengers (see Table 4 and Figure 9). Not surprisingly, the cost per passenger rises when the study area widens and if the VoT increases. It is instead more interesting to evaluate the "gain" in cost that we can achieve via *Adaptive Transit*, with respect to the *MRT-FRF* scheme:

$$Gain = \frac{Z^{24h}_{\text{MRT-FRF}} - Z^{24h}_{\text{Adaptive}}}{Z^{24h}_{\text{MRT-FRF}}} \quad (23)$$

where $Z^{24h}_{\text{MRT-FRF}}$ and $Z^{24h}_{\text{Adaptive}}$ are the respective optimal costs computed as in Equation 21.

From Figure 9a, we notice that the gain of *Adaptive Transit* is consistent in all the scenarios. To better interpret the result, we also represent agency-related (Figure 9b) and user-related cost (Figure 9c) separately. With low VoT (10 €/h) or small $R$ (10 km) *Adaptive Transit* brings a modest increase of agency-related costs (~2%), but a significant reduction of user-related costs (from 6% to 8%), which brings to a significant improvement of total cost. When considering higher values of VoT and $R$ (VoT = 20 €/h and $R > 30$ km), instead, the optimal configuration of the *Adaptive Transit* leads to an increase of agency-related costs (~30%), which is larger than the decrease in user-related costs (~12%). One reason is that *Adaptive Transit*, as already claimed, prevents access to MRT from degrading in suburban areas, by offering a demand-responsive feeder service with relatively high frequency. This may be inefficient when the areas to be served are too big (big *R*). Moreover, when VoT is very high, *Adaptive Transit* is inefficient in terms of agency-cost, in order to accommodate well users to minimize overall cost.

*Table 4. Cost gains (on daily basis) of the Adaptive scheme with respect to the MRT-FRF scheme for different combinations of city radius (R) and Value of Time ($\mu_A$).*

|  |  | Total cost gain (%) | | | Agency-related cost gain (%) | | | User-related cost gain (%) | | |
|---|---|---|---|---|---|---|---|---|---|---|
| R (km) | VoT (€/h) | 10 | 15 | 20 | 10 | 15 | 20 | 10 | 15 | 20 |
| 10 |  | 4.5 | 4.7 | 4.2 | -2.0 | -1.4 | -1.8 | 7.0 | 6.8 | 6.1 |
| 15 |  | 4.8 | 3.1 | 3.7 | -1.7 | -10.7 | -6.5 | 7.2 | 7.4 | 6.6 |
| 20 |  | 5.0 | 3.3 | 3.2 | -1.8 | -11.9 | -13.3 | 7.5 | 8.0 | 7.6 |
| 25 |  | 5.0 | 3.6 | 2.8 | -1.9 | -12.9 | -24.1 | 7.7 | 8.7 | 10.1 |
| 30 |  | 5.1 | 2.7 | 2.6 | -2.1 | -19.5 | -29.3 | 7.9 | 9.8 | 11.6 |
| 35 |  | 5.0 | 2.4 | 2.3 | -2.2 | -22.0 | -31.6 | 7.9 | 10.4 | 12.1 |
| 40 |  | 5.0 | 1.6 | 1.5 | -2.5 | -24.2 | -32.6 | 8.0 | 10.2 | 11.8 |

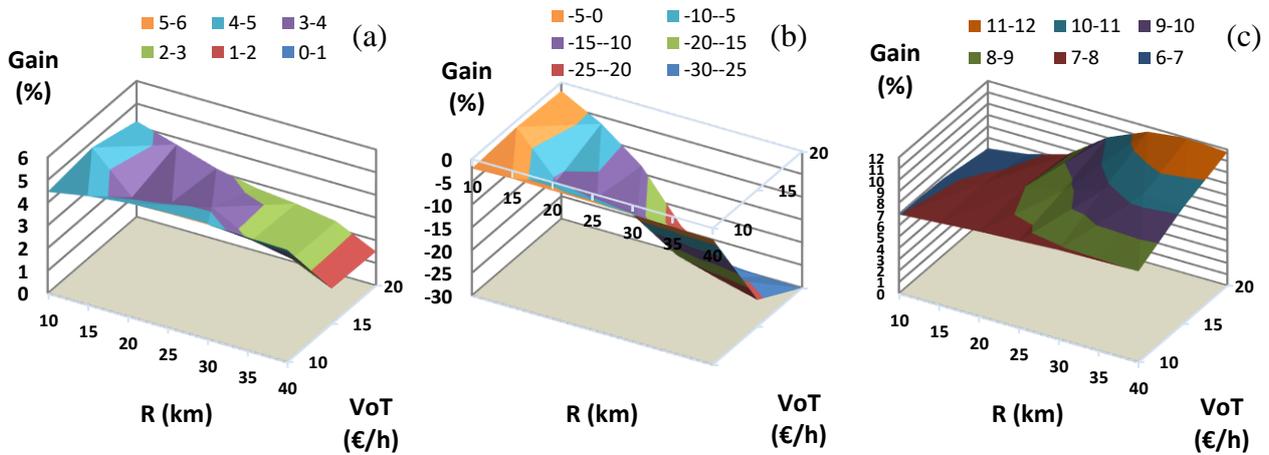

*Figure 9. Cost gain (on daily basis) of the Adaptive scheme with respect to the MRT-FRF scheme for different combinations of city radius (R) and Value of Time ($\mu_A$), considering: (a) the total cost; (b) the agency-related costs; (c) the user-related costs (for result interpretation, the colours represent the different value ranges of the cost gain).*

Overall, these results confirm that *Adaptive Transit* is always more cost-effective than *MRT-FRF* in the different scenarios considered. The additional agency-cost are largely compensated by the reduction of user-cost.

## 5.6 Automated vehicles scenario

In this section, we explore a future scenario where the transit services (both MRT and feeder) are provided via automated vehicles, which are expected to decrease operating cost. With respect to the non-automated scenario, we consider a reduced cruising speed ($v_{FRF}, v_{DRF}$) of 18 km/h for the feeder buses (see Tirachini and Antoniou (2020) in Section 5.1) and a dwell time per passenger ($\tau_p$) of 3 s. We assume that the agency can save 50% of crew salaries, and not 100% since back-office personnel and new safety devices in vehicles will be needed (Tirachini and Antoniou, 2020). Therefore, we set $\mu_{M,MRT}$ = 100 €/veh h and $\mu_{M,FRF} = \mu_{M,DRF}$ = 30 €/veh h (instead of the values in Table 3). While this section is a first attempt to evaluate the benefits of *Adaptive Transit* with automated vehicles, future work should analyse in more detail other important variables (e.g., willingness of passengers to use them) and operation conditions (e.g., road infrastructure, commercial speed, etc.), as well as the ratio between the drivers' salary and the vehicle cost (see Tirachini and Antoniou (2020) – Tables 1 and 2).

Figure 10 shows that *Adaptive Transit* tilts the balance towards the reduction of user-related cost, more noticeably with automation than without (walking cost $Z_A$ reduced by 35% with respect to the *MRT-FRF* scheme, against the 27% reduction in the non-automated case). This is expected, since the operational cost reduction brought by automation allows the agency to deploy a more user-centric service. This however results in increased capital cost to acquire more vehicles for that service (capital cost $Z_{cap}$ increases by 32% with respect to the *MRT-FRF* scheme in the automated case, against the 13% in the non-automated case).

Comparing Figure 10 (automated case) with Figure 9 (non-automated case) we see that during peak hours DRF is deployed at shorter distance from the city centre and the headway of the feeder service is smaller. No remarkable differences appear during the off-peak.

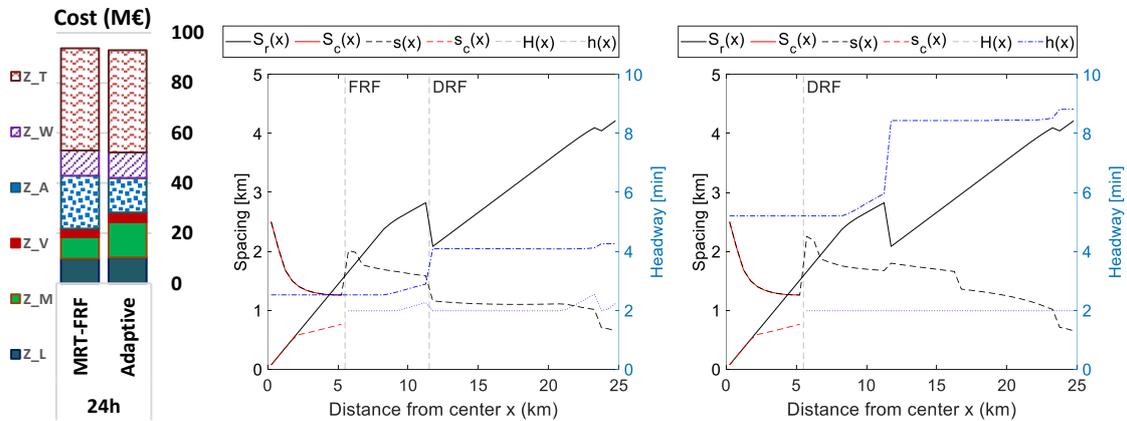

*Figure 10. Results for the automated case: Total daily cost (left) and optimal decision variables for the Adaptive Transit scheme in the peak (centre) and off-peak (right).*

# 6. Conclusion and future research

We have presented the concept of *Adaptive Transit*, which combines fixed-route and demand-responsive transportation and alternates between the two services in order to adapt to the spatial and temporal variation of the demand density. We provided a theoretical high-level model of *Adaptive Transit* based on Continuous Approximation, where the demand density and the decision variables defining the transit network are continuous functions across the space and vary over time. Numerical results on such a model show that *Adaptive Transit* tilts the balance of the overall costs in favour of user-centric components, while keeping the agency-related cost at a reasonable level, such that the total cost (the sum of the two) is improved. In particular, with respect to conventional transit (*MRT-FRF* scheme), *Adaptive Transit*:

- Improves the user-related cost (-8.7%) but requires higher agency-related costs (+1.1%). The overall total cost is reduced (-3.6%). The cost savings are more pronounced during the off-peak (-7.2%) – see Section 5.2. These trends emerge even more clearly assuming automated vehicles (see Section 5.6).
- Provides particular benefits to the far suburbs, reducing access time to MRT stations (-32.1% during the off-peak) – see Section 5.3.
- Is particularly advantageous when the size of the study area and the users' VoT are relatively small (total cost reduction between 4% and 5%) – see Section 5.5.

An additional advantage, not considered in this work, may be the induced or latent demand due to the modal shift from car to transit, which would take place thanks to the improved service for passengers.

To summarize, the novelty of this work is that it provides managerial insights on how to holistically optimize the design of a future transit system, able to combine fixed schedule and demand-responsive operations in a single multi-modal service. In future work, we will systematically verify the impact of electrification and automation, which deeply modify the agency-related cost components. Also, we will extend the model considering more complex urban settings, such as a polycentric one. We believe the concept of *Adaptive Transit* can guide planning agencies in the deployment of more efficient next-generation transit systems.

## Acknowledgments

This research was supported by the National Research Foundation under its CREATE program and the Singapore-MIT Alliance for Research and Technology, Future Urban Mobility Interdisciplinary Research Group. The work has been partially supported by the project "WEAKI-TRANSIT: WEAK-demand areas Innovative TRANsport Shared services for Italian Towns" (unique project code: E44I17000050001) under the Italian programme "PRIN 2017", by the project "ADDRESS" under the University of Catania programme "PIACERI Linea 2", by the French ANR research project MuTAS (ANR-21-CE22-0025-01) and by the research grant, funded by Institut Francais (Ambassade de France en Italie), received by G. Calabrò.

# Appendix A. Derivation of the cost components and constraints of the model

The derivation of the cost components and constraints of the Continuous Approximation (CA) model is presented as follows. The abbreviations used throughout the paper are listed in Table A1. Table A2 lists the parameters and variables used in the model.

*Table A1. Acronyms used in this article*

| | |
|---|---|
| CA | Continuous Approximation |
| DR | Demand Responsive |
| DRF | Demand Responsive Feeder |
| FMLM | First Mile and Last Mile |
| FR | Fixed-Route |
| FRF | Fixed-Route Feeder |
| MRT | Mass Rapid Transit |
| PT | Public Transit |
| PVT | Private Vehicle Trips |
| QoS | Quality of Service |
| VKT | Vehicle-Kilometres Travelled |

*Table A2. Notation of the CA model's variables and parameters*

| | |
|---|---|
| **Independent variables** | |
| $x \in [0, R]$ | Radial distance from the centre ($x = 0$) of the urban area |
| $t \in \mathcal{T}$ | Time of day. |
| **Input parameters** | |
| $R$ | Radius of the metropolitan area |
| $\rho_0(t)$ | Demand density in the city centre during the time slot $t$ |
| $\rho(x)$ | Demand density at distance $x$ from the city centre (pax/km$^2$ h) (Equation 5) |
| $\gamma$ | Slope of the Clark's law (demand density distribution - Equation 5) |
| $v_{MRT}, v_{FRF}, v_{DRF}, v_w$ | Cruise speed of MRT, FRF, DRF; walking speed |
| $C_{MRT}, C_{FRF}, C_{DRF}$ | Vehicle capacity |
| $\tau_{s,MRT}, \tau_{s,FRF}, \tau_{s,DRF}$ | Dwell time at MRT stations, FRF and DRF stops |
| $\tau_p$ | Extra dwell time per passenger |
| $\mu_A, \mu_W, \mu_T$ | *VoT* of the access, walking and in-vehicle travel time |
| $\delta_{tr}$ | Time penalty due to transfers |
| $\mu_{L,MRT}, \mu_{L;FRF}, \mu_{L,DRF}$ | Cost coefficients related to the infrastructure length |
| $\mu_{V,MRT}, \mu_{V;FRF}, \mu_{V,DRF}$ | Cost coefficients for the distance travelled by the vehicles |
| $\mu_{M,MRT}, \mu_{M;FRF}, \mu_{M,DRF}$ | Cost coefficients related to the fleet size |
| $N(x)$ | Number of passengers in the infinitesimal annulus of radius $x$, $N(x) = 2\,\rho(x) \cdot (2\pi x)$ |
| **Local decision variables at location $x$** | |
| $\theta_r(x), S_r(x)$ | Angular and linear spacing between radial MRT lines |
| $S_c(x)$ | Spacing between ring MRT lines |
| $s(x)$ | Spacing between stations along a radial MRT line |
| $\phi(x)$, $s_c(x)$ | Angular and linear spacing between stations on a ring MRT line |
| $H(x)$ | Headway on ring and radial MRT lines |
| $h(x)$ | Headway of the feeder service in the suburban area |
| $d_j(x)$ | Spacing between FRF stops if $j$ = FRF, twice the maximum walking distance from the station if $j$ = DRF |
| $\mathcal{F}(x)$ | Type of FMLM service, $\mathcal{F}(x) \in \{FRF, DRF, 0\}$ |
| $N_s(x)$ | Number of FMLM strips. |
| **Derived variables at location x** | |
| $\mathbb{I}_j(x)$ | Indicator function, it is 1 if $\mathcal{F}(x) = j$, 0 otherwise |
| $D(x), D$ | Local decision functions $D(x) = \{\theta_r(x), S_c(x), s(x), \phi(x), H(x), h(x), d_j(x), \mathcal{F}(x)\}$ $D = \{D(x)\}_{x \in [0,R]}$ |
| $l(x)$ | Length of the FMLM rectangle (Equation 6) |
| $w(x)$ | Width of one strip within the FMLM rectangle (Equation 6) |
| $CL_j(x)$ | Cycle length of the FMLM feeder of type $j \in \{FRF, DRF\}$ |
| $C_j(x)$ | Cycle time of the FMLM feeder of type $j \in \{FRF, DRF\}$ |
| **Global decision variables** | |
| $r$ | Radius of the central (double-coverage) area |
| $Q_0$ | Maximum value of total radial flow of MRT vehicles, $Q_0 = Q(x=0)$ |
| $H_B$ | Headway at the outermost ring line $H_B = H(x = r)$ |
| $G$ | Global decision variables $G = (r; Q_0; H_B)$ |
| **Output parameters** | |
| $y_{L,j}(x), y_{V,j}(x), y_{M,j}(x)$ | Agency local cost at distance $x$ from the centre: infrastructure, vehicle-km and fleet costs, $j \in \{MRT, FRF, DRF\}$ |
| $y_{A,j}(x), y_{W,j}(x), y_{T,j}(x)$ | Travel time components suffered by users at distance $x$ from the city centre: walking, waiting and in-vehicle travel time, $j \in \{MRT, FRF, DRF\}$ |
| $F_L, F_V, F_A, F_W, F_T$ | Global cost components for agency and users. |
| $Dem_o(x), Dem_d(x), Dem(x)$ | Number of trips originated (or destined) within rings of radii $(x, x + dx)$. We assume $Dem_o(x) = Dem_d(x) = Dem(x)$ |
| $P_o(x), P_d(x), P(x)$ | Probability for a trip to have origin (or destination) lying within rings of radii $(x, x + dx)$. We assume $P_o(x) = P_d(x) = P(x)$ (Equation A.1) |
| $DEM$ | Total number of trips per hour in the study area (see Equation A.1). |
| $v_{cr}(x), v_{cc}(x)$ | Commercial speed on radial lines and ring lines at $x$ (see Equation A.2). |
| $Q(x)$ | Radial flow of MRT vehicles (trains per hour) |
| $O_j(x)$ | Average vehicle occupancy, $j \in \{MRT, DRF, FRF\}$ |
| $Z(D,G,t)$ | Total cost at time $t$ (see Equation 16) |
| $Z^{24h}$ | Total cost, over the entire day (see Equation 21) |

The derivation procedure (in Appendices A1-A3) of the cost components for the MRT is based on Chen et al. (2015). We repeat the formulas, for clarity, but the reader who is familiar with the calculus in Chen et al. (2015) can skip them. The novel computation we need to model *Adaptive Transit* is concentrated in Appendices A4-A6. Also observe that, differently from Chen et al. (2015), we consider the same headway on ring and radial lines $H_r(x) = H_c(x) = H(x)$ and treat the spacing between radial $s(x)$ and ring $\phi(x)$ MRT stations as local decision variables, while in Chen et al. (2015) they do not change with $x$. Also, in this study we consider a symmetric demand pattern, so that $Dem_o(x) = Dem_d(x) = Dem(x) = 2\pi x \cdot \rho(x)$, and we obtain:

$$P(x) = \frac{Dem(x)}{DEM} = \frac{Dem(x)}{\int_0^R Dem(y)dy} \tag{A.1}$$

As explained in Section 4.5, agency's and user's metrics are converted into cost density functions by means of a set of cost coefficients, in order to compute the total cost (per unit of time) as a linear combination of those metrics. We denote with $\mu_{L,MRT}$ or $\mu_{L,FMLM}$ (€/km-h), $\mu_{V,MRT}$ or $\mu_{V,FMLM}$ (€/veh-km) and $\mu_{M,MRT}$ or $\mu_{M,FMLM}$ (€/veh-h) the cost coefficients related to the agency metrics, for MRT and FMLM, respectively. Regarding the MRT, we also consider costs specifically related to the stations through a coefficient $\mu_{ST}$ (€/station).

## A.1 Derivation of operational outputs and constraints for the MRT

<u>MRT radial-line and ring-line commercial speed</u>. Along the radial MRT line, the time needed to travel 1 unit of distance is $1/v_{MRT}$ plus the time spent at each station $\tau_{s,MRT}$ for acceleration, deceleration, and boarding/alighting, multiplied by the number $\frac{1}{s(x)}$ of stations in the unit of distance. Similar reasoning can be applied to MRT ring lines. The resulting speed on ring and radial lines is, respectively:

$$v_{cc}(x) = 1/\left(\frac{1}{v_{MRT}} + \frac{\tau_{s,MRT}}{s(x)}\right); \quad v_{cr}(x) = 1/\left(\frac{1}{v_{MRT}} + \frac{\tau_{s,MRT}}{\phi(x)\cdot x}\right) \tag{A.2}$$

<u>Commercial speed on the boundary ring</u>. As before, but considering the angle $\phi_B$ between stations on the outermost ring line, the speed on the boundary line ($x = r$) is as follows:

$$v_{cB} = 1/\left(\frac{1}{v_{MRT}} + \frac{\tau_{s,MRT}}{\phi_B \cdot r}\right) \tag{A.3}$$

<u>Vehicle capacity constraint</u>. The expected maximum number of passengers on board a MRT vehicle is constrained to be less than the vehicle's passenger-carrying capacity, i.e., $O_{MRT}(x) \leq C_{pax,MRT}$. To better understand the derivation of the following formulas, the reader can refer to the scheme of Figure A1. Based on Section 4.3, passengers using ring lines are approximately those whose angle between origin and destination is less than 2 radians. The percentage of such passengers is $2 \cdot \frac{2\text{ rad}}{2\pi} = 2/\pi$. Therefore, for ring lines, the total number of passengers onboard all MRT vehicles on a ring line at $x$ is: $P_o(x) \int_x^R P_d(y)dy \cdot \frac{2}{\pi} + P_d(x) \int_x^R P_o(y)dy \cdot \frac{2}{\pi} = 2\left(P(x) \int_x^R P(y)dy\right) \cdot \frac{2}{\pi}$. Since the travelled angular distance along a ring line is comprised between 0 and 2 radians, the average travelled distance is 1 radian, i.e., x km of linear travelled distance. Therefore, the ratio of the average trip length over the length of the ring line is: $x/(2\pi x) = 1/(2\pi)$. The flow of MRT vehicles through that ring in each direction is: $1/(S_c(x)H(x))$, which means $2/(S_c(x)H(x))$ considering both directions. Hence (see Appendix 15 of Chen et al. (2015)):

$$O_{MRT,c}(x) = 2\left(Dem(x)\int_x^R P(y)dy\right) \cdot \frac{2}{\pi} \cdot \frac{S_c(x)H(x)}{2} \cdot \frac{1}{2\pi} = \left(Dem(x)\int_x^R P(y)dy\right) \cdot \frac{2}{\pi} \cdot \frac{S_c(x)H(x)}{2\pi},$$

$$\text{if } x < r. \quad (A.4)$$

Observe that for $x > r$ there are no ring MRT lines.

As for radial lines, note that the users travelling along radial MRT lines crossing ring at $x < r$ are:

- The users departing from $y > x$ and arriving at $y < x$ and whose angle between origin and destination is less than 2 radians. The percentage of such users is $\int_x^R P_o(y)dy \int_0^x P_d(y)dy \cdot \frac{2}{\pi}$.
- The users departing from $y < x$ and arriving at $y > x$ and whose angle between origin and destination is less than 2 radians. The percentage of such users is $\int_x^R P_d(y)dy \int_0^x P_o(y)dy \cdot \frac{2}{\pi}$.
- The users departing from $y > x$ and whose angle between origin and destination is more than 2 radians (independent of the ring where the destination is).
- The users arriving to $y > x$ and whose angle between origin and destination is more than 2 radians (independent of the ring where the origin is).

Therefore, we obtain (see Appendix A15 of Chen et al. (2015)):

$O_{MRT,r}(x)$
$= DEM \cdot \left(\int_x^R P_o(y)dy \int_0^x P_d(y)dy \cdot \frac{2}{\pi} + \int_x^R P_o(y)dy \cdot \left(1 - \frac{2}{\pi}\right) + \int_x^R P_d(y)dy \int_0^x P_o(y)dy \cdot \frac{2}{\pi} + \int_x^R P_d(y)dy \cdot \left(1 - \frac{2}{\pi}\right)\right) \cdot \frac{\theta_r(x)H(x)}{2} \cdot \frac{1}{2\pi}$

$= DEM \cdot \left(\int_x^R P(y)dy \int_0^x P(y)dy \cdot \frac{2}{\pi} + \int_x^R P(y)dy \cdot \left(1 - \frac{2}{\pi}\right)\right) \cdot \frac{\theta_r(x)H(x)}{2\pi}, \quad \text{if } x < r. \quad (A.5)$

Instead, users travelling along radial MRT lines crossing ring at $x > r$ are only those departing or arriving at distance $y > x$:

$$O_{MRT,r}(x) = DEM \cdot \left(\int_x^R P_o(y)dy + \int_x^R P_d(y)dy\right) \cdot \frac{S_c(x)H(x)}{2} \cdot \frac{1}{2\pi} = DEM \cdot \left(\int_x^R P(y)dy\right) \cdot \frac{S_c(x)H(x)}{2\pi}$$

$$\text{if } x > r. \quad (A.6)$$

<u>Vehicle capacity constraint on the boundary ring.</u> Similarly, the vehicle's passenger-carrying capacity constraint for the boundary MRT line is: $O_B \leq C_{pax,MRT}$. Hence:

$$O_B = DEM \cdot \left(\int_x^R P(y)dy \cdot \int_x^R P(y)dy\right) \cdot \frac{2}{\pi} \cdot \frac{H_B/2}{2\pi}. \quad (A.7)$$

**A.2 Derivation of the agency-related cost components for the MRT**

The cost components related to the transit agency depend on:

- Infrastructure length $L$ (km), i.e., the construction and maintenance costs. For the MRT we also include an additional unit cost component related to stations.
- Total distance $V$ (veh km/h) travelled by the vehicles per unit of time, i.e., the operational costs.
- Size $M$ (veh) of the vehicle fleet, i.e., the capital cost for vehicle ownership and crew costs.

<u>Local cost for the length of MRT lines</u>. Consider the area between two rings of radii $x$ and $x + dx$, which is equal to $2\pi x \cdot dx$. The length of the MRT radial lines within the ring pair is $2\pi/\theta_r(x) \cdot dx$. The length of the ring lines is instead $2\pi x/S_c(x) \cdot dx$ (recall that they only exist in $x \leq r$). The local cost is given by the length of the MRT lines in the area divided by the area width $dx$; that is:

$$Y_L(x) = \frac{2\pi}{\theta_r(x)} + \frac{2\pi x}{S_c(x)} \text{ if } x < r; \quad Y_L(x) = \frac{2\pi}{\theta_r(x)}, \text{ if } x > r.$$

<u>Local cost for the vehicle-distance travelled per hour</u>. It is obtained by multiplying the local cost for the length of ring-lines and radial-lines with their corresponding transit flows, i.e., $1/H(x)$, multiplied by 2 due to the bi-directional travel flows on each line; that is:

$$Y_V(x) = \frac{4\pi}{\theta_r(x)H(x)} + \frac{4\pi x}{S_c(x)H(x)} \text{ if } x < r; \quad Y_V(x) = \frac{4\pi}{\theta_r(x)H(x)}, \text{ if } x > r.$$

<u>Global cost for the vehicle-distance travelled per hour</u>. Global costs refer only to the outermost (boundary) ring line, hence:

$$F_V = \frac{4\pi r}{H_B}.$$

<u>Local cost for the fleet size</u>. It is given by the ratio between the vehicle-distance travelled per hour and the commercial speed, that is (see Appendix A3 of Chen et al. (2015):

$$Y_M(x) = \frac{4\pi}{\theta_r(x)H(x)v_{cr}(x)} + \frac{4\pi x}{S_c(x)H(x)v_{cc}(x)} \text{ if } x < r; \quad Y_M(x) = \frac{4\pi}{\theta_r(x)H(x)v_{cr}(x)}, \text{ if } x > r.$$

<u>Global cost for the fleet size</u>. As before:

$$F_M = \frac{4\pi r}{H_B v c B}.$$

Note that $Y_M(x)$ and $F_M$ are calculated considering the decision variable values of the peak hours. We do so, as the agency must dimension its fleet to peak hours.

**A.3 Derivation of the user-related cost components for the MRT**

The cost components related to the users of the transit system depend on:

- Walking time $A$ to reach the bus stop or the MRT station or the destination.

- Waiting time $W$ at the bus stop or the MRT station.
- In-vehicle travel time $T$ including boarding, riding, dwell and alighting time.
- Transfers: since any possible transfer between different transit lines is an additional disutility, we treat it as a penalty of the extra walking time $\Delta A$.

To derive the user-related cost components, we consider nine different cases of trip changes, as depicted in Figure A1. Observe that these cases depend on where origin and destination are (suburbs or centre) and the angle Θ between origin and destination.

Global cost for the transfers between MRT lines.

This computation is based on the one of Chen et al. (2015), appendix A.14.

We assume the convention in the literature (e.g., Badia et al., 2014) and we assume that during their trips travellers aim to minimize the distance. In order to do so, they:

- Transfer once when either the origin or the destination is in the centre (cases (b),(c),(e),(f)) and when both origin and destination are in the periphery and the angle between them is $\Theta \geq 2$ rad (case (a)).
- Transfer twice when origin and destination both lie in the periphery and the angle between them $\Theta < 2$ rad (case (d))
- Do not transfer, when the closest line to both origin and destination is the same, which happens when $\Theta_1 \leq \theta_{r,min} = \min_{x \in [0,R]} \theta_r(x)$ or when the closest ring line between origin and destination is the same, which happens when $|x - y| < S_c(x)/2$, where $x$ and $y$ are the rings where the origin and the destination lie, respectively. The cases without transfer are (g)-(i).

Observe that the probability for any trip to have no transfers is $\Pr(\text{no transfers}) = \frac{\theta_{r,min}}{2\pi}$. The probability to have both origin and destination in the periphery is $\int_r^R P_0(y) \cdot dy \cdot \int_r^R P_d(y) \cdot dy$ and the probability for the angle between them to be $\theta < 2$ rad is $\frac{2}{\pi}$. Therefore, the probability for any trip to have two transfers is $\Pr(\text{2 transfers}) = \left(\frac{2}{\pi} - \frac{\theta_{r,min}}{2\pi}\right) \cdot \int_r^R P_0(y) \cdot dy \cdot \int_r^R P_d(y) \cdot dy$. Therefore, by simple calculation[13], the average expected number of transfers can be approximated by:

$$F_A = 1 + \int_r^R P(y)dy \cdot \int_r^R P(y)dy \cdot \left(\frac{2}{\pi} - \frac{\theta_{r,min}}{\pi}\right).$$

Local cost for the passenger average walk time to/from MRT.

This computation is based on the one of Chen et al. (2015), appendix A.8.

---

[13] Our calculation is similar to Appendix A.14 of Chen et al. (2015), except that we also allow travellers to have no transfers (see Figure 3a and the related comment).

Cases (a) and (g): $Y_A(x) = 2 \cdot Dem(x) \int_r^R P(y)dy \cdot \left(\frac{\theta_r(x)x}{4} + \frac{s}{4}\right)/v_w \cdot \left(1 - \frac{2}{\pi}\right)$, if $x > r$.

Cases (b) and (h): $Y_A(x) = \begin{cases} 2 \cdot Dem(x) \int_r^R P(y)dy \cdot \left(\frac{\theta_r(x)x}{4} + \frac{s}{4}\right)/v_w \cdot \left(1 - \frac{2}{\pi}\right), & \text{if } x < r \\ 2 \cdot Dem(x) \int_0^r P(y)dy \cdot \left(\frac{\theta_r(x)x}{4} + \frac{s}{4}\right)/v_w \cdot \left(1 - \frac{2}{\pi}\right), & \text{if } x > r. \end{cases}$

Cases (c) and (i): $Y_A(x) = 2 \cdot Dem(x) \int_0^r P(y)dy \cdot \left(\frac{\theta_r(x)x}{4} + \frac{s}{4}\right)/v_w \cdot \left(1 - \frac{2}{\pi}\right)$, if $x < r$.

Case (d): $Y_A(x) = 2 \cdot Dem(x) \int_r^R P(y)dy \cdot \left(\frac{\theta_r(x)x}{4} + \frac{s}{4}\right)/v_w \cdot \frac{2}{\pi}$, if $x > r$.

Cases (e): $Y_A(x) = \begin{cases} 2 \cdot Dem(x) \int_r^R P(y)dy \cdot \left(\frac{\phi(x)x}{4} + \frac{S_c}{4}\right)/v_w \cdot \frac{2}{\pi}, & \text{if } x < r \\ 2 \cdot Dem(x) \int_0^r P(y)dy \cdot \left(\frac{\theta_r(x)x}{4} + \frac{s}{4}\right)/v_w \cdot \frac{2}{\pi}, & \text{if } x > r. \end{cases}$

Case (f): $Y_A(x) = 2 \cdot Dem(x) \int_x^r P(y)dy \cdot \left(\frac{\phi(x)x}{4} + \frac{S_c}{4}\right)/v_w \cdot \frac{2}{\pi} + 2 \cdot Dem(x) \int_0^x P(y)dy \cdot \left(\frac{\theta_r(x)x}{4} + \frac{s}{4}\right)/v_w \cdot \frac{2}{\pi}$, if $x < r$.

If we sum up these components (Cases (a)-(i)) we obtain:

$Y_A(x) = \begin{cases} 2 \cdot Dem(x) \cdot \left[\int_0^x P(y)dy \cdot \left(\frac{\theta_r(x)x}{4} + \frac{s}{4}\right)/v_w + \int_x^R P(y)dy \cdot \left(\frac{\phi(x)x}{4} + \frac{S_c}{4}\right)/v_w\right], & \text{if } x < r \\ 2 \cdot Dem(x) \cdot \left(\frac{\theta_r(x)x}{4} + \frac{s}{4}\right)/v_w, & \text{if } x > r. \end{cases}$

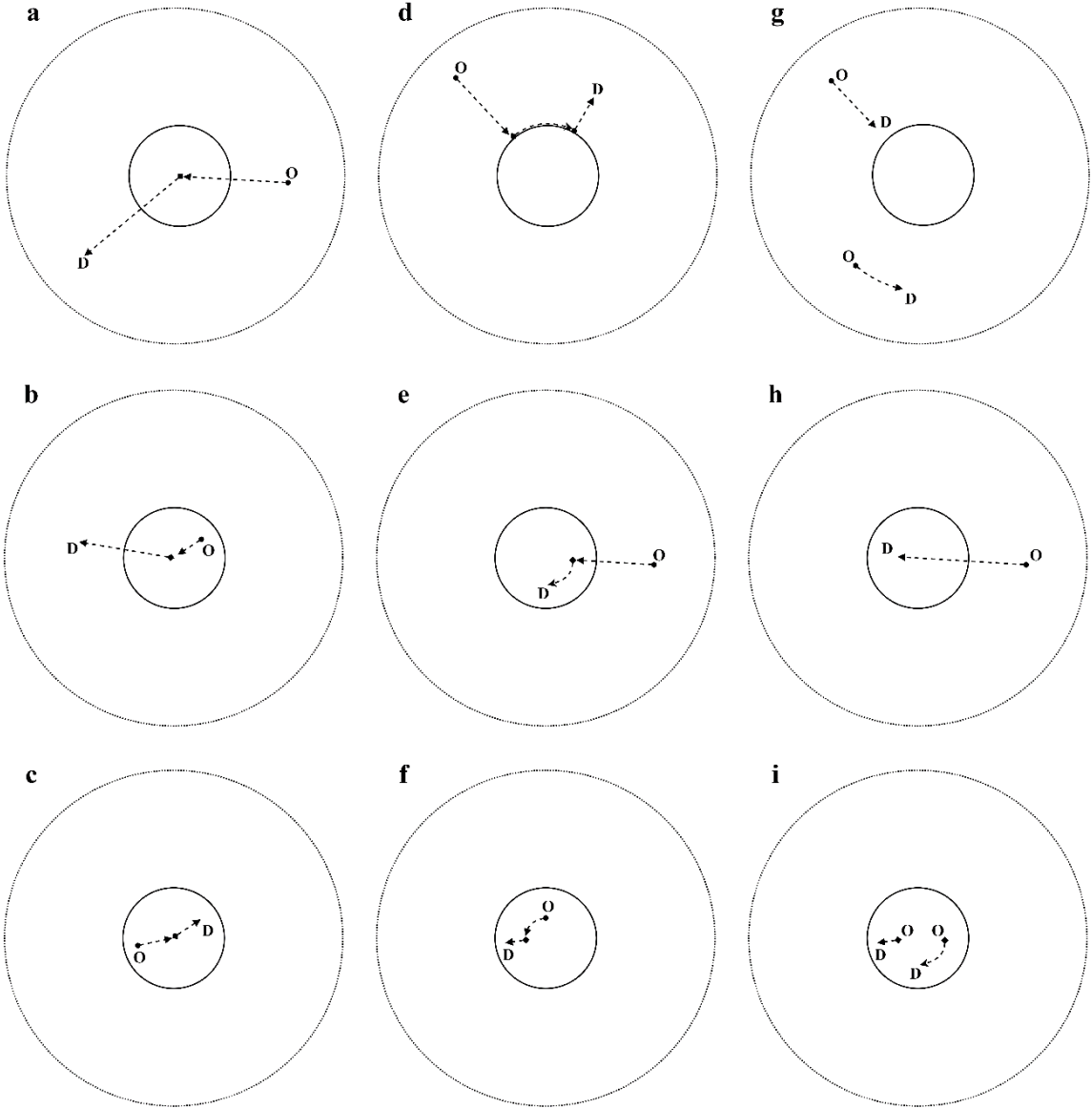

*Figure A1. Different cases of trips for the derivation of user-related cost components.*

Local cost for the passenger average waiting time.

This computation is based on the one of Chen et al. (2015), appendix A.6.

Case (a): $Y_W(x) = 2 \cdot Dem(x) \int_r^R P(y)dy \cdot \frac{H(x)}{2} \cdot \left(1 - \frac{2}{\pi}\right)$, if $x > r$.

Case (b): $Y_W(x) = \begin{cases} 2 \cdot Dem(x) \int_r^R P(y)dy \cdot \frac{H(x)}{2} \cdot \left(1 - \frac{2}{\pi}\right), \text{ if } x < r \\ 2 \cdot Dem(x) \int_0^r P(y)dy \cdot \frac{H(x)}{2} \cdot \left(1 - \frac{2}{\pi}\right), \text{ if } x > r. \end{cases}$

Case (c): $Y_W(x) = 2 \cdot Dem(x) \int_0^r P(y)dy \cdot \frac{H(x)}{2} \cdot \left(1 - \frac{2}{\pi}\right)$, if $x < r$.

Case (d): $Y_W(x) = 2 \cdot Dem(x) \int_r^R P(y)dy \cdot \frac{H(x)}{2} \cdot \frac{2}{\pi}$, if $x > r$.

Case (e): $Y_W(x) = \begin{cases} 2 \cdot Dem(x) \int_r^R P(y)dy \cdot \frac{H(x)}{2} \cdot \frac{2}{\pi}, & \text{if } x < r \\ 2 \cdot Dem(x) \int_0^r P(y)dy \cdot \frac{H(x)}{2} \cdot \frac{2}{\pi}, & \text{if } x > r. \end{cases}$

Case (f): $Y_W(x) = 2 \cdot Dem(x) \int_0^r P(y)dy \cdot \frac{H(x)}{2} \cdot \frac{2}{\pi}$, if $x < r$.

The following cases account for those trips which do not require transfers between MRT lines. Therefore, they should be subtracted from the total value of $Y_W(x)$.

Cases (g), (h) and (i): $Y_W(x) = \begin{cases} 2 \cdot Dem(x) \int_0^x P(y)dy \cdot \frac{H(x)}{2} \cdot \frac{\theta_r(x)}{\pi}, & \text{if } x < r \\ 2 \cdot Dem(x) \int_{x-s/2}^{x+s/2} P(y)dy \cdot \frac{H(x)}{2} \cdot \frac{\theta_r(x)}{\pi}, & \text{if } x > r. \end{cases}$

If we sum up Cases (a-g) and subtract Cases (g-i) we obtain:

$Y_W(x) = \begin{cases} 2 \cdot Dem(x) \cdot \left[1 - \left(\int_0^x P(y)dy\right) \cdot \frac{\theta_r(x)}{\pi}\right] \cdot \frac{H(x)}{2}, & \text{if } x < r \\ 2 \cdot Dem(x) \cdot \left[1 - \left(\int_0^x P(y)dy + \int_{x-s/2}^{x+s/2} P(y)dy\right) \cdot \frac{\theta_r(x)}{\pi}\right] \cdot \frac{H(x)}{2}, & \text{if } x > r. \end{cases}$

<u>Global cost for the passenger average waiting time</u>.

This computation is based on the one of Chen et al. (2015), appendix A.7.

We only consider the cost that occurs at the outermost ring line, for case (d), excluding the limited portion of trips which do not involve any transfer:

$F_W = DEM \cdot \int_r^R P(y)dy \cdot \int_r^R P(y)dy \cdot \frac{H_B}{2} \cdot \left(\frac{2}{\pi} - \frac{\theta_{r,min}}{\pi}\right).$

<u>Local cost for the passenger average in-vehicle travel time</u>.

This computation is based on the one of Chen et al. (2015), appendix A.12.

We consider the annulus $(x, x + dx)$. We obtain the in-vehicle travel time on each annulus weighted by the proportion of users who travel along it and who cross it. Then, the in-vehicle travel time on that annulus by its width $dx$.

Cases (a), (b) and (c): $Y_T(x) = DEM \cdot 2 \int_x^R P(y)dy \cdot \frac{1}{v_{cr}(x)} \cdot \left(1 - \frac{2}{\pi}\right), \forall x$

Case (d): $Y_T(x) = DEM \cdot 2 \int_x^R P(y)dy \cdot \frac{1}{v_{cr}(x)} \cdot \frac{2}{\pi}$, if $x > r$.

Cases (e) and (f): $Y_T(x) = DEM \cdot 2 \int_x^R P(y)dy \cdot \int_0^x P(y)dy \cdot \frac{1}{v_{cr}(x)} \cdot \frac{2}{\pi} + 2 \cdot Dem(x) \int_x^R P(y)dy \cdot \frac{x}{v_{cc}(x)} \cdot \frac{2}{\pi}$, if $x < r$

The following cases account for those trips which do not require transfers between MRT lines. Therefore, they should be subtracted from the total value of $y_{T,MRT}(x)$.

Case (g): $Y_T(x) = DEM \cdot 2 \int_{x+\frac{s}{2}}^{x+\frac{s}{2}} P(y)dy \cdot \int_{x+\frac{s}{2}}^{x+\frac{s}{2}} P(y)dy \cdot \frac{1}{v_{cr}(x)} \cdot \frac{\theta_r(x)}{\pi}$, if $x > r$.

Case (h): $Y_T(x) = DEM \cdot 2 \int_x^R P(y)dy \cdot \int_0^x P(y)dy \cdot \frac{1}{v_{cr}(x)} \cdot \frac{\theta_r(x)}{\pi}$, if $x > r$.

If we sum up Cases (a-g) and subtract Cases (g-h) we obtain the local cost $Y_T(x)$.

Global cost for the passenger average in-vehicle travel time.

This computation is based on the one of Chen et al. (2015), appendix A.13.

We consider the in-vehicle travel time along the boundary ring line:

$F_T = DEM \cdot \int_r^R P(y)dy \cdot \int_r^R P(y)dy \cdot \frac{r}{v_{cB}} \cdot \left(\frac{2}{\pi} - \frac{\theta_{r,min}}{\pi}\right)$.

**A.4 Derivation of the agency-related cost components for the FMLM**

We now compute the agency-related cost for FMLM. Recall that it is defined only for $x > r$. The number of lines crossing the ring at $x$ is $2\pi/\theta_r(x)$. At any $x$, each line is adjacent to two FLML areas (see Figure 2). Therefore, the number of FLML sub-regions along the ring at $x$ is $sa(x) = 2\pi/(\theta_r(x)/2)$. Each FMLM sub-region is divided in $N_s(x)$ strips and has a width of $s(x)$. We assume that the infrastructure cost at distance $x$ does not depend on whether FRF or DFR is deployed there. Therefore, considering the length of the two-way FRF bus infrastructure ($CL_{FRF}(x)/2$, as in Equation 7):

$$y_L(x) = N_s(x) \cdot \frac{CL_{FRF}(x)/2}{s(x)} \cdot sa(x)$$

We need to do a distinction for $y_V(x)$ and $y_M(x)$, in which the cycle length $CL_j(x)$ and the cycle time $C_j(x)$ appear, which are different for FRF and DRF (see Equations 7;9;12;13). The vehicle-distance travelled per hour is obtained by multiplying the cycle length, per unit of distance, $CL_j(x)/s(x)$ by their corresponding vehicle frequency $1/h(x)$:

$$y_V(x) = \sum_{j=\{FRF,DRF\}} N_s(x) \cdot \frac{CL_j(x)}{s(x) \cdot h(x)} \cdot sa(x) \cdot \mathbb{I}_{FRF}(x)$$

The cost of fleet size is derived from the number of vehicles $Cj/(s(x) \cdot h(x))$ needed to ensure the feeder service:

$$y_M(x) = \sum_{j=\{FRF,DRF\}} N_s(x) \cdot \frac{C_j(x)}{s(x) \cdot h(x)} \cdot sa(x)$$

In the equations above, *sa(x)* multiplies on the left the cost (per distance unit) related to a single service sub-region.

## A.5 Derivation of the user-related cost components for the FMLM

Let us denote with the walking speed with $v_w$ and the number of passengers in the infinitesimal annulus of radius $x$ with:

$$N(x) = 2\,\rho(x) \cdot (2\pi x)$$

Note that a percentage $p_{walk}(x) = p_{\{walk,FRF\}}(x) \cdot \mathbb{I}_{FRF}(x) + p_{\{walk,DRF\}}(x) \cdot \mathbb{I}_{DRF}(x)$ (see equations 8;11) of passengers does not use the feeder service as they reside in the walking area (so they prefer to walk directly to MRT). Then, the total time components suffered by users at $x$ in the FMLM are expressed as follows. FRF passengers walk to the closest stop (feeder stop or directly MRT station). According to Equation 8, the average walk distance is $s(x)/4 + d(x)/4$. DRF passengers are picked-up and dropped-off in-place, without walking. Only the users within the walking area (the grey triangle in Figure 4), i.e., a fraction $p_{walk,DRF}(x)$, walk to/from the MRT station an average distance of $(2/3)\,d_{0,DRF}(x)$. This value is derived computing the mean of the horizontal and vertical distance from the MRT station, weighted by the demand density. Therefore, in the $x$ where a feeder is deployed, the local cost for walking to access the feeder (or directly MRT) is:

$$y_{A,j}(x) = \begin{cases} N(x) \cdot \dfrac{s(x)+d(x)}{4\,v_w} & \text{and} \quad \text{if } j = \text{FRF} \\ N(x) \cdot p_{walk,DRF}(x) \cdot \dfrac{2/3\,d_{0,DRF}(x)}{4\,v_w} & \text{if } j = \text{DRF} \end{cases}$$

The average waiting time for the feeder service and the in-vehicle travel time inside the feeder bus are as follows. Note that they are not experienced by passengers directly walking to the MRT station, whence the term $(1 - p_{walk,j}(x))$. Assuming that the average waiting time for both feeder services is given by the half of the headway $h(x)$, the local cost for the feeder waiting time is:

$$y_{W,j}(x) = N(x) \cdot \left(1 - p_{walk,j}(x)\right) \cdot \dfrac{h(x)}{2} \quad \text{if } j \in \{FRF, DRF\}$$

The average in-vehicle travel time for the FRF case is given by $C_{FRF}/4 + (\Delta l(x) + d(x)/2)/v_{FRF}$. We recall that $\Delta l(x)$ is the average extra vertical distance (see Figure 4) which the FRF has to travel due to the different position of the strips with respect to the MRT station they serve, that we approximate to $s(x)/4$ if $N_s(x) > 1$, and 0, otherwise. For the DRF case, the average in-vehicle travel time is given by $C_{DRF}/4 + \Delta l(x)/v_{DRF}$. The local cost for the feeder in-vehicle travel time is:

$$y_{T,j}(x) = \begin{cases} N(x) \cdot \left(1 - p_{walk,FRF}\right) \cdot \left[\dfrac{C_{FRF}}{4} + \dfrac{\left(\Delta l(x) + \frac{d(x)}{2}\right)}{v_{FRF}}\right] & \text{if } j = \text{FRF} \\ N(x) \cdot \left(1 - p_{walk,DRF}\right) \cdot \left[\dfrac{C_{DRF}}{4} + \dfrac{\Delta l(x)}{v_{DRF}}\right] & \text{if } j = \text{DRF} \end{cases}$$

We can summarize the formulas above, for any FMLM feeder type, by using indicator functions.

$$y_A(x) = \sum_{j \in \{FRF, DRF\}} \mathbb{I}_j(x) \cdot y_{A,j}(x)$$

$$y_W(x) = \sum_{j \in \{FRF, DRF\}} \mathbb{I}_j(x) \cdot y_{W,j}(x)$$

$$y_T(x) = \sum_{j \in \{FRF, DRF\}} \mathbb{I}_j(x) \cdot y_{T,j}(x).$$

**A.6 Derivation operational constraints for the FMLM**

<u>Vehicle capacity constraint</u>. The expected maximum number of passengers on board a FRF or a DRF vehicle is constrained to be less than the vehicle's passenger-carrying capacity, i.e., $O_j(x) \leq C_{pax,j}, j \in \{FRF, DRF\}$. The demand originated in one unit of time within a strip of a FMLM sub-region is $\rho(x) \cdot w(x) \cdot l(x)$. Of these passengers, only a fraction $1 - p_{\text{walk},j}(x)$ takes the feeder bus. The number of buses in a unit of time is $\frac{1}{h(x)}$. Therefore, the number of passengers per feeder vehicle is

$$O_j(x) = \rho(x) \cdot w(x) \cdot l(x) \cdot (1 - p_{\text{walk},j}(x)) \cdot h(x), \text{ if } x > r. \tag{A.8}$$

# Appendix B. Agent-based simulation of a small-scale scenario

The methodology adopted throughout this paper is Continuous Approximation (CA). CA allows obtaining managerial insights on the impact of different design parameters on the overall performance of transit and the trade-offs involved. Therefore, it is a useful high-level tool for transportation planners. However, the CA model of transit is abstract and it is not easy to translate it to a real deployment. In reality, transit lines cannot be as symmetric as assumed in our CA model. Moreover, the assumption of representing demand via a density function is highly idealized, since the real demand results from the behaviour of different individual travellers.

To study the benefits of *Adaptive Transit* removing such idealizations, we resort in this appendix to Agent-Based Modelling. However, studying *Adaptive Transit* in the same large-scale scenarios targeted with CA would require huge computational effort, which is out of the scope of this paper and would deserve a separate article. We instead focus on a small case scenario, in which we confirm that the concept of *Adaptive Transit*, i.e., appropriately shifting bus operating regimen between fixed-route and demand-responsive, greatly reduces user-related costs, while preventing agency-related cost from exploding.

**B.1 Simulation scenarios**

We simulate three hypothetical scenarios (labelled as SC0, SC1, SC2) on a prototypical urban network as follows:

- Scenario **SC0** represents the baseline scenario involving fixed route (FR) transit and private modes; in particular, the following modes are available to travellers: private modes (including car, car-pooling, taxi) and public transit, i.e., FR bus or MRT or the combination of the two, with walk access.
- In scenario **SC1**, there are no FR buses. Transit users access MRT either by walking or via demand-responsive feeder (DRF) buses, directly from their doorsteps.
- In scenario **SC2**, *Adaptive Transit* is introduced, which operates the FR buses during the peak (when higher capacity vehicles are required) and DRF services during the off-peak. In other words, SC2 is equal to SC0 during the peak and to SC1 during the off-peak.

We utilize an agent-based traffic simulation platform (called SimMobility Mid-term) that models individual daily activity patterns through an activity-based model (ABM) system (Ben-Akiva et al., 1996) and simulates the trajectories of vehicles and passengers on a multimodal network using a mesoscopic traffic simulator (combining the speed density relationships with a queuing model). Details may be found in Adnan et al. (2016), Lu et al. (2015). Recently, SimMobility has been enhanced with a controller that models the fleet operations of demand-responsive services (i.e., Shared AV taxi (Oh et al., 2020a), Minibus service (Oh et al., 2020b)).

Observe that in this appendix we do not want to mimic exactly the scenarios of the main paper (Section 3.2), as in any case a strict comparison between the results of CA and simulation would not be useful, due to the enormous difference in the scale of the scenarios considered. This appendix has thus to be considered as a separate small-scale study to confirm, from another angle, the potential benefits of the concept of *Adaptive Transit*.

**B.2 Simulation settings**

The three scenarios are compared in a prototypical urban network, which is moderately sized and consists of land use patterns and household socio-demographics that resemble Singapore in small scale. The synthetic population consists of 350,000 individuals distributed over the geographical region shown in Figure B1a. Details of the synthesis of this prototypical network are available in Zhu and Ferreira (2014) and Basu and Ferreira (2021). The network consists of 95 nodes connected by 254 links across 24 traffic analysis zones. There are 12 bus lines covering 86 bus stops, and 2 MRT lines traversing 20 stations. The central business district (CBD) is also demarcated on the map.

Figure B1b shows the demand pattern over the time-of-day, which we assume fixed across scenarios, which involves around 532,000 trips for the 24 hour period. The mode shares in the baseline scenario (SC0) are: 14.7% (bus), 19% (MRT), 21.6% (private vehicle trips, PVT), and 44.8% (Walk from origin to destination).

After preliminary empirical parameter value optimization, the fleet size for the DRF service is set to be 1000 vehicles, which ensures 100% of satisfaction rate of trip requests within a waiting time threshold of 10 mins (See more details on fleet sizing in Oh et al. 2020a, 2020b). FR bus and MRT operate with a predefined dispatching headway: 5.2 min (06:30 - 08:30 for AM, 17:00 - 20:30 for PM) during the peak and 12.7 min during the off-peak on average, which are aligned with the Google transit network data in Singapore. This results in 642 and 820 bus departures from the terminal during the peak/off-peak respectively. Also note that there are a total of 120 departures for each MRT line during a day. Since the focus of this study is on the performance of transit, the share between private modes and transit is kept fixed along the three scenarios.

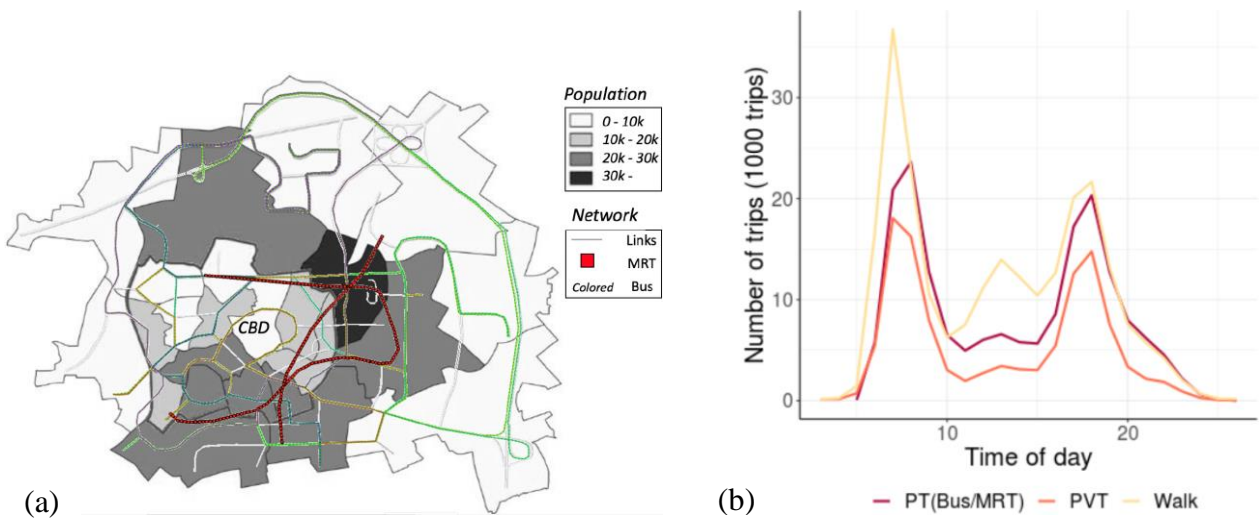

(a) (b)

*Figure B1. Prototypical urban network and demand pattern: (a) Road and transit network; (b) Demand distribution.*

### B.3 Simulation results

The three scenarios clearly result in different operating costs which can be approximated by VKT (Vehicle-Km Travelled), shown in Figure B2. The VKT generated by the DR service in SC1 and SC2 include both 'operational' trips (e.g., empty trips for cruising, parking, pick-up) and 'service' trips from pick-up to drop-off points. Private trips (PVT) are the predominantly responsible for VKT (generating more than 545,000 veh-km), while Transit only accounts for 7% of VKT in SC0. However, the footprint of transit becomes very high, up to 44% in SC1, due to the DR vehicles. Moreover, most of such VKT is "lost" in deadheading. Integrating FR and DR, as in SC2, brings instead the transit footprint back to 22%. Scenario SC2 reduces significantly the total VKT of feeder by more than 70% compared to that of SC1. To understand how SC2 achieves such a reduction, Figure B2b shows the VKT of the DR fleet by time-of-day, and indicates that in SC1 a significant portion of VKT is generated during the peak periods. This is expected, as the DR service is incapable of efficiently serving high demand rates and would result in "tortuous" DR vehicle trajectories (Araldo et al., 2019), longer service/operational distance (Oh et al., 2020b), and network congestion (Oh et al. 2020a; Oh et al. 2021). Such an evident increase of VKT is prevented in SC2, during peak, by removing DRF buses and adopting fixed buses.

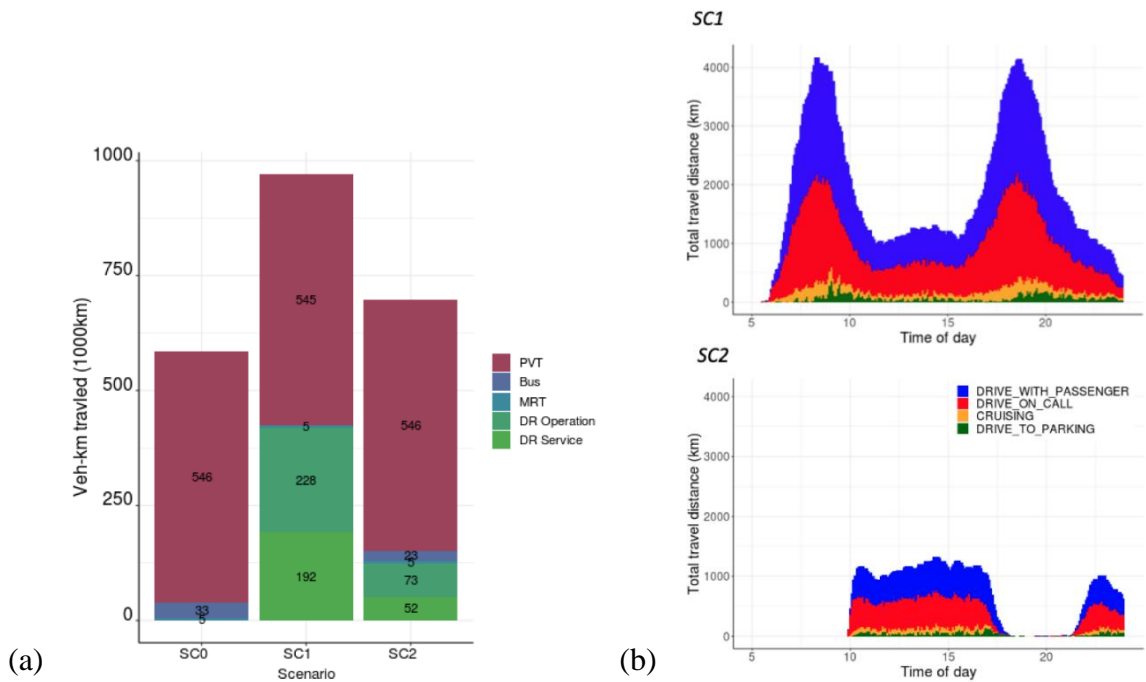

*Figure B2. Vehicle-Km Travelled: (a) Overall VKT; (b) VKT of DR (SC1, SC2).*

Despite the operational savings described above, SC2 does not degrade the travel time of users, as it is visible in Table B1, in which we report the average travel time components of users, depending on the kind of trip they choose, i.e., unimodal or multimodal.

Observe that in SC0 the lower frequency of FR bus service during off-peak causes waiting time to be more than double than during peak hours. Instead, in SC1, the waiting time for the bus (DRF in that case) is sufficiently low, but worsens during the peak (waiting time for bus in SC1 is more than double than in SC0), as the DR vehicle routes are not able to efficiently serve high demand.

Combining FR and DR buses, as in scenario SC2, provides instead the best balance, as the waiting time is as good as SC0 during peak, and as good as SC1 during off-peak[14]

*Table B1. Average of travel time components of transit users (unit: minutes)*

| Period | SC | Trip | Pax (10³) | Walk time | DR feeder bus | | FR bus | | MRT | |
|---|---|---|---|---|---|---|---|---|---|---|
| | | | | | *wait* | *in-veh* | *wait* | *in-veh* | *wait* | *in-veh* |
| Peak | SC0 | Multimodal | 1.3 | 6.6 | - | - | 2.9 | 6.7 | 4.0 | 4.5 |
| | | Unimodal: Bus | 101.2 | 6.7 | - | - | 3.3 | 8.7 | - | - |
| | | Unimodal: MRT | 36.1 | | - | - | - | - | 4.0 | 4.4 |
| | SC1 | Multimodal | 35.4 | 0 | 5.9 | 6.6 | - | - | 4.6 | 4.7 |
| | | Unimodal: MRT | 70.3 | 10.5 | - | - | - | - | 4.0 | 4.7 |
| | SC2 | Multimodal | 4.1 | 6.7 | - | - | 3.1 | 6.6 | 4.0 | 4.5 |
| | | Unimodal: Bus | 97.0 | 6.7 | - | - | 3.2 | 8.5 | - | - |
| | | Unimodal: MRT | 44.5 | | | | - | - | 4.0 | 4.5 |
| Off Peak | SC0 | Multimodal | 3.1 | 6.6 | - | - | 6.8 | 6.3 | 4.6 | 4.1 |
| | | Unimodal: Bus | 54.8 | 6.9 | - | - | 6.6 | 7.4 | - | - |
| | | Unimodal: MRT | 14.7 | | - | - | - | - | 4.6 | 4.1 |
| | SC1 | Multimodal | 16.5 | 0 | 5.6 | 5.7 | - | - | 5.0 | 4.6 |
| | | Unimodal: MRT | 42.9 | 10.4 | - | - | - | - | 4.4 | 4.2 |
| | SC2 | Multimodal | 12.5 | 0 | 5.5 | 5.6 | - | - | 4.5 | 4.6 |
| | | Unimodal: MRT | 10.8 | 10.8 | - | - | - | - | 4.5 | 4.6 |

To recap, the FR bus, as in SC0, is inefficient during off-peak hours (high waiting time). On the other hand, the DRF, as in SC1, is inefficient during the peak (high operational cost). *Adaptive Transit* can get the best of FR buses and DRF (in terms of both agency and user related cost) by shifting between them, as in SC2, depending on time of day.

---

[14] One might be tempted to expect that the results concerning SC0 and SC2 to be the same during peak hours. Instead, the table shows that they are not exactly equal. This is due to the stochasticity of travel behaviors and network dynamics in simulation. The same occurs for SC1 and SC2 during off-peak. All simulation scenarios were achieved through several iterations to minimize this variability (Oh et al. 2021).